\documentstyle[12pt]{article}
\def\dir{}

\def\figcond{1}

\def\figsty{0}

\def\authname{}

\newcount\figsubcountss \figsubcountss=0
\ifnum\figcond>0
\input epsf
  \ifnum\figsty>0
  \fi
\fi
\def\figinsert#1#2#3#4{
\ifnum\figcond>0
  \ifnum\figsty>0
    \ifnum\figsubcountss=0
      \immediate\write9{\noexpand\catcode`\noexpand\@=11}
      \immediate\write9{\noexpand\newpage}
      \immediate\write9{\noexpand\pagestyle {empty}}
      \immediate\write9{\noexpand\section* {Figure Captions}}
      \immediate\write9{\noexpand\begin {enumerate}}
      \immediate\write9{\noexpand\renewcommand {\noexpand\theenumi}{}}
      \immediate\write9{\noexpand\begin {enumerate}}
      \immediate\write9{\noexpand\renewcommand
        {\noexpand\theenumii}{\noexpand\arabic {enumii}}}
      \immediate\write9{\noexpand\renewcommand {\noexpand\labelenumii}
        {Fig. \noexpand\arabic {enumii}:}}
      \immediate\write10{\noexpand\newpage}
      \ifnum\figsty>1
        \immediate\write10{\noexpand\pagestyle {headings}}
        \immediate\write10{\noexpand\setcounter {page}{1}}
        \immediate\write10{\noexpand\renewcommand {\noexpand\thepage}
            {Figure \noexpand\arabic{page} -- \noexpand\authname}}
      \else
        \immediate\write10{\noexpand\pagestyle {empty}}
      \fi
    \fi
    \global\advance\figsubcountss by 1
    \immediate\write10{\noexpand\begin {figure}[p]}
    \immediate\write10{\noexpand\begin {center}}
    \immediate\write10{\noexpand\ }
    \immediate\write10{\noexpand\epsfbox {\dir #1}}
    \immediate\write10{\noexpand\ \noexpand\\}
    \ifnum\figsty<2
      \immediate\write10{\noexpand\vspace {1cm}}
      \immediate\write10{Figure \noexpand\ref {#3}}
    \fi
    \immediate\write10{\noexpand\end {center}}
    \immediate\write10{\noexpand\end {figure}}
    \immediate\write10{\noexpand\newpage}
   \let\save=\ref \let\ref=0 \let\savec=\cite \let\cite=0
    \immediate\write9{\noexpand\item #2}
   \let\ref=\save \let\cite=\savec
    \immediate\write9{\noexpand\label {#3}}
    \begin {figure}[htbp]
    \begin{center}
    \fbox{Fig. \ref{#3}}
    \end{center}
    \end {figure}
  \else
    \begin{figure}[#4]
    \begin{center}
    \ \epsfbox{\dir #1}
    \caption []{#2 \label {#3}}
    \end{center}
    \end{figure}
  \fi
\else
  \begin {figure}[htbp]
  \begin{center}
  \fbox{Fig. \ref{#3}}
  \caption []{#2 \label {#3}}
  \end{center}
  \end {figure}
\fi
}
\def\Closeout#1{%
   \immediate\closeout#1}
\def\figepsfout{
\Closeout10
  \immediate\write9{\noexpand\end {enumerate}}
  \immediate\write9{\noexpand\end {enumerate}}
  \immediate\write9{\noexpand\catcode`\noexpand\@=12}
\Closeout9
\input \jobname.cap
\input \jobname.fis}
\catcode`\@=11
\newbox\tempboxa
\newdimen\captionboxsubcount
\def\capsize#1{\captionboxsubcount=#1pt}
\newdimen\captionboxsub
\captionboxsub=\hsize \advance\captionboxsub by -\captionboxsubcount
\advance\captionboxsub by -\captionboxsubcount
\long\def\@makecaption#1#2{
 \setbox\@tempboxa\hbox{#1: #2}
 \ifdim \wd\@tempboxa >\captionboxsub
\rightskip=\captionboxsubcount \leftskip=\captionboxsubcount #1: #2
\else \hbox to\hsize{\hfil\box\@tempboxa\hfil}
 \fi}
\def\enddocument{
\ifnum\figcond>0
  \ifnum \figsty>0 \figepsfout
\fi\fi
\@checkend{document}\clearpage\begingroup
\if@filesw \immediate\closeout\@mainaux
\def\global\@namedef##1##2{}\def\newlabel{\@testdef r}%
\def\bibcite{\@testdef b}\@tempswafalse \makeatletter\input \jobname.aux
\if@tempswa \@warning{Label(s) may have changed.  Rerun to get
cross-references right}\fi\fi\endgroup\deadcycles\z@\@@end}
\catcode`\@=12
\capsize{30}

\renewcommand{\theequation}{\thesection.\arabic{equation}}

\renewcommand{\today}{\ifcase\month\or
 Jan.\or Feb.\or Mar.\or Apr.\or May\or Jun.\or
 Jul.\or Aug.\or Sep.\or Oct.\or Nov.\or Dec.\fi
 \space\number\day, \number\year}

\begin{document}


\thispagestyle{empty}
\setcounter{page}{0}
\baselineskip=22pt

\begin{titlepage}
\begin{flushright}
\begin{tabular}{l}
SU-4240-624\\
hep-ph/9511335\\
November 1995
\end{tabular}
\end{flushright}
\bigskip
\begin{center}
\LARGE
\bf
Simple~Description~of~$\pi\pi$~Scattering~to~1~GeV
\end{center}
\vfill
\begin{center}
\Large
{\it Masayasu Harada $^{a}$\footnote{{\it e-mail:}~mharada@npac.syr.edu}},
{}~~~~~~~~ {\it Francesco Sannino $^{a,b}$ \footnote{{\it e-mail:}
 ~sannino@npac.syr.edu}}
\vskip .4cm
{\large \it and}
\vskip .4cm
{\it
Joseph Schechter $^{a}$ \footnote{{\it e-mail:}~schechter@suhep.phy.syr.edu}}
\end{center}
\vfill
\begin{itemize}
\large\it
\smallskip
\item[$^a$] {Department of Physics, Syracuse University, Syracuse,
New York,
13244-1130.}
\item[$^b$] {Di\-par\-ti\-mento di Scienze Fi\-si\-che, Mo\-stra
d'Ol\-tre\-mare
Pad.\-19, 80125
Na\-po\-li, Ita\-lia.}
\end{itemize}
\vfill
\begin{abstract}
Motivated by the $\displaystyle{1/N_c}$ expansion, we present a simple model of
$\pi\pi$ scattering as a sum of a {\it current-algebra} contact term
and resonant pole exchanges. The model preserves crossing symmetry as
well as unitarity up to $1.2~GeV$. Key features include chiral
dynamics, vector meson dominance, a broad low energy scalar ($\sigma$)
meson and a {\it Ramsauer-Townsend} mechanism for the
understanding of the $980~MeV$ region. We discuss in detail the {\it
regularization} (corresponding to rescattering effects) necessary to
make all these nice features work.
\vfill
\end{abstract}
\vfill
\end{titlepage}
\newpage

\section{Introduction}
\setcounter{equation}{0}
Historically, the analysis of $\pi\pi$ scattering has been considered an
important test of our understanding of strong interaction physics
(QCD, now) at low energies. It is commonly accepted that the key feature is
the approximate spontaneous breaking of chiral symmetry. Of course, the {\it
kinematical} requirements of unitarity and crossing symmetry
should be respected. The chiral perturbation scheme \cite{chp}, which improves
the tree
Lagrangian approach by including loop corrections and counterterms, can
provide a description of the scattering up to the energy region slightly
above threshold $(400-500~MeV)$.

In order to describe the scattering up to energies beyond this region (say
to around $1~GeV$) it is clear that the effects of particles lying in this
region must be included and some new principle invoked. A plausible hint
comes from the large $N_c$ approximation to QCD, in which the leading
order scattering amplitudes consist of just tree diagrams containing resonance
exchanges as well as possible contact diagrams \cite{1n}. The method suggests
that an infinite number of resonances are required and also a connection
with some kind of string theory \cite{string}.

Some encouraging features were previously found in an approach which
truncated the particles appearing in the effective Lagrangian  to those with
masses up to an energy slightly greater than the range of interest. This
seems reasonable phenomenologically and is what one usually does in
setting up an effective Lagrangian. The most famous example is the chiral
Lagrangian of only pions. In  Ref.~\cite{Sannino-Schechter} this Lagrangian
provided, as a starting point, a contact term which described the threshold
region. However the usual observation was made that the real part of the
$I=0,~J=0$ partial wave amplitude quite soon violated the unitarity bound
$|R^0_0|\leq 1/2$ rather severely. The inclusion of the contribution
coming from the $\rho$ meson exchange was observed to greatly improve,
although not completely solve, this problem. These results are shown
explicitly in Fig.~\ref{figura1} and provide some encouragement for the
possible success of a truncation scheme.

\figinsert{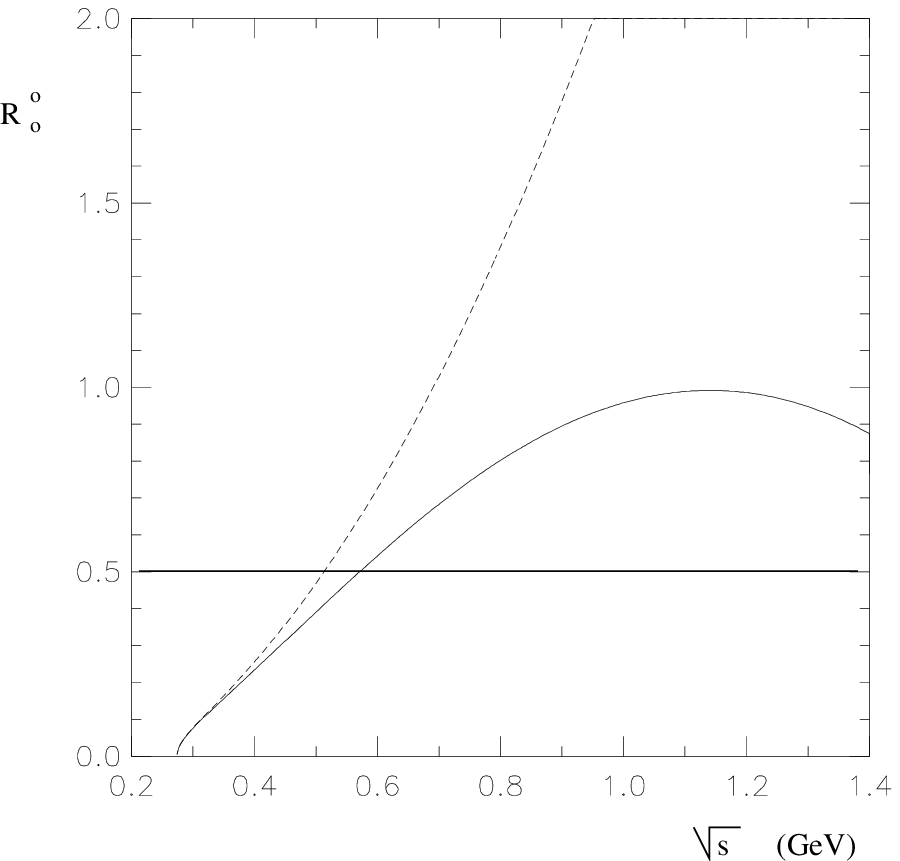}{Predicted curves for $R^0_0$. The solid line
which shows the {\it current algebra}
$+$ $\rho$ result for $R^0_0$ is much closer to the unitarity bound of
$0.5$ than the dashed line which shows the {\it current
algebra} result alone. }{figura1}{hptb}

In  Ref.~\cite{Sannino-Schechter}, it was observed that the inclusion of
resonances
up till and including the $p$-$wave$ region enabled one to construct an
amplitude which
satisfied the unitarity bounds up to about $1.3~GeV$. It was assumed
that, above this point, new resonances would come in to preserve
unitarity. This hypothesis was called {\it local cancellation}.
The model produced a reasonable looking
$I=J=0$ phase shift up to about $800~MeV$. In this paper we will attempt
to describe and carefully compare with experiment the interesting physics
lying between $800$ and $1200~MeV$ in this truncated $\displaystyle{1/N_c}$
inspired framework.
Specifically we will be concerned with the proper inclusion of the $f_0(980)$
scalar resonance as well as the opening of the $K\overline{K}$ channel.
We find that a simple reasonable description of the $f_0(980)$ region is
obtained when the interplay of this resonance with its background is
taken into account. In this approach the background amplitude is predicted
by the model itself.
In the region just above the $K\overline{K}$ threshold we notice the
feature analogous to the elastic case that the severe unitarity violation of
the inelastic
$\pi\pi\rightarrow K\overline{K}$ amplitude is damped by the inclusion of
vector
meson and scalar meson exchange diagrams.

Of course, it would be wonderful if one could simply add the various
contributions to the tree level amplitude and find a good match to
experiment. This is not possible for a variety of reasons, which are
discussed in Section 2. The needed {\it regularizations} are
introduced there. Section 3 gives a brief overview of the model and
reviews the important role of a broad scalar meson in the low energy
($<800~MeV$) region. Section 4 contains a discussion of various
aspects of the $1~GeV$ region. The characteristic feature - a type of
{\it Ramsauer-Townsend} effect resulting from the interplay of the
$f_0(980)$ resonance with the predicted background - is outlined in
section 4.1 and treated in more detail in 4.2. In section 4.3 it is
shown that the introduction of the {\it next group} of resonances,
located in the $1300~MeV$ region, does not make major changes in the
$\pi\pi$ scattering below $1200~MeV$ (the changes are essentially
absorbed in small changes of the parameters of the broad low energy
scalar). In section 4.4 it is demonstrated that the
phenomenological introduction of inelastic effects associated with
the opening of the $K\overline{K}$ channel does not make a significant
change in our picture of $\pi\pi \rightarrow \pi\pi$ below $1200~MeV$.
Section 4.5 contains a presentation of the $I=J=0$ phase shift
obtained by combining our predicted real part with unitarity. In
section 5 we discuss the inelastic $\pi\pi \rightarrow K \overline{K}$
channel and show that here also the resonance exchanges damp the
unitarity bound violation due to the contact term. Section 6 contains
the summary and further discussion. Finally, Appendices~A, B and C give
details on, respectively, the scattering kinematics, the chiral
Lagrangian and the unregularized amplitudes.

\section{Difficulties of the Approach}
\setcounter{equation}{0}
In the large $N_c$ picture the leading amplitude (of order
$\displaystyle{1/N_c}$) is a sum
of polynomial contact terms and tree type resonance exchanges. Furthermore
the resonances should be of the simple $q\overline{q}$ type; glueball and
multi-quark
meson resonances are suppressed. In our phenomenological model there is
no way of knowing {\it a priori} whether a given experimental state is
actually of $q\overline{q}$ type. For definiteness we will keep all relevant
resonances even though the status of a low lying scalar resonance like
the $f_0(980)$ has been considered especially controversial
\cite{Tornqvist:95}. If such
resonances turn out
in the future to be not of $q\overline{q}$ type, their tree contributions would
be
of higher order than $\displaystyle{1/N_c}$. In this event the amplitude would
still of course
satisfy crossing symmetry.

The most problematic feature involved in comparing the leading
$\displaystyle{1/N_c}$ amplitude
with experiment is that it does not satisfy unitarity. In fact, resonance
poles like
\begin{equation}
\frac{1}{M^2-s}
\label{propagator}
\end{equation}
will yield a purely real amplitude, except at the singularity, where
they will diverge and drastically violate the unitarity bound. Thus in order
to compare the $\displaystyle{1/N_c}$ amplitude with experiment we must
regularize the denominators in some way. The usual method, as
employed in Ref.~\cite{Sannino-Schechter}, is to regularize the propagator so
that the resulting partial wave amplitude has the  locally unitary form
\begin{equation}
\frac{M\Gamma}{M^2-s-iM\Gamma}\ .
\label{Breit-Wigner}
\end{equation}
This is only valid for a narrow resonance in a region where the
{\it background} is negligible. Note that the $-iM\Gamma$ is strictly
speaking a higher order in $\displaystyle{1/N_c}$ effect.

For a very broad resonance
there is no guarantee that such a form is correct. Actually, in
Ref.~\cite{Sannino-Schechter} it was found necessary to include a rather broad
low lying scalar resonance (denoted $\sigma(550)$) to avoid violating the
unitarity bound. A suitable form turned out to be of the type
\begin{equation}
\frac{M G}{M^2-s-iMG^\prime}\ ,
\label{sigma-propagator}
\end{equation}
where $G$ is not equal to the parameter $G^\prime$ which was
introduced to regularize the propagator. Here $G$ is the quantity
related to the squared coupling constant.

Even if the resonance is narrow, the effect
of the background may be rather important. This seems to be true for the
case of the $f_0(980)$. Demanding local unitarity in this case yields a partial
wave amplitude of the well known form \cite{Taylor}:
\begin{equation}
\frac{e^{2i\delta}M\Gamma}{M^2-s-iM\Gamma}+e^{i\delta}\sin \delta\ ,
\label{rescattering}
\end{equation}
where $\delta$ is a background phase (assumed to be slowly varying). We
will adopt a point of view in which this form is regarded as a kind of
regularization of our model. Of course, non zero
$\delta$ represents a rescattering
effect which is of higher order in $\displaystyle{1/N_c}$. The quantity
$\displaystyle{e^{2i\delta}}$, taking $\delta=constant$, can be
incorporated into the squared
coupling constant connecting the resonance to two
pions. In this way, crossing symmetry can be preserved. {}From its origin, it
is clear that the complex residue does not signify the existence of a
{\it ghost} particle. The non-pole background term in
eq.~(\ref{rescattering}) and hence $\delta$
is to be predicted by the other pieces in the effective Lagrangian.

Another point which must be addressed in comparing the leading
$\displaystyle{1/N_c}$ amplitude
with experiment is that it is purely real away from the singularities. The
regularizations mentioned above do introduce some imaginary pieces but these
are clearly more model dependent. Thus it seems reasonable to compare the
real part of our predicted amplitude with the real part of the experimental
amplitude. Note that the difficulties mentioned above arise only for
the direct channel poles; the crossed channel poles and contact terms will
give purely real finite contributions.

It should be noted that if we predict the real part of the amplitude, the
imaginary part can always be recovered by assuming elastic unitarity
(which is likely to be a reasonable approximation up to about $1~GeV$).
Specializing eq.~(\ref{real-imaginary}) in Appendix~A to the $\pi \pi$
channel we have for the imaginary piece $I^I_l$ of the $I,~l$ partial
wave amplitude
\begin{equation}
I^I_l=\frac{1}{2}\left[1\pm\sqrt{{\eta^I_l}^2-4 {R^I_l}^2}\right]\ ,
\label{imaginary}
\end{equation}
where $\eta^I_l$ is the elasticity parameter. Obviously this formula
is only meaningful if the real part obeys the bound
\begin{equation}
|R^I_l| \leq \frac{\eta^I_l}{2}\ .
\label{real-bound}
\end{equation}
The main difficulty one has to overcome in obtaining a unitary amplitude
by the present method is the satisfaction of this bound. Therefore, one
sees that making {\it regularizations} like eqs.~(\ref{Breit-Wigner}) and
(\ref{rescattering}) which provide
unitarity in the immediate region of a narrow resonance is not at all
tantamount to unitarizing the model by hand. One might glance again
at Fig.~\ref{figura1} for emphasis of this point.

To summarize this discussion, we will proceed by comparing the real part
of a suitably regularized tree amplitude computed from a chiral Lagrangian of
pseudoscalar
mesons and resonances with the real part of the experimental amplitude
deduced  from the standard phase shift analysis.

\section{Overview and Low Energy Region}
\setcounter{equation}{0}

The amplitude will be constructed from the non-linear chiral Lagrangian
briefly summarized in Appendix~B. To start with, we shall neglect the
existence of the $K$ mesons. Then the form of the unregularized
amplitude is identical to the one presented  in Ref.~\cite{Sannino-Schechter}.
The neutral resonances which can contribute have the quantum numbers
$J^{PC}=0^{++},~1^{--},$ and $ 2^{++}$. We show in Table 1
\noindent
\begin{table}
\begin{center}
\begin{tabular}{|c||c|| c c c|} \hline \hline
  &$I^G(J^{PC})$ & $M(MeV)$ & $\Gamma_{tot}(MeV)$ & $Br(2\pi)\%$\\ \hline
\hline
$\sigma(550)$ & $0^+(0^{++})$  & 559     & 370      & $-$  \\
$\rho(770)$   & $1^+(1^{--})$  & 769.9   & 151.2    & 100  \\
$f_0(980)$   & $0^+(0^{++})$  & 980     & 40$-$400 & 78.1 \\
$f_2(1270)$  & $0^+(2^{++})$  & 1275    & 185      & 84.9 \\
$f_0(1300)$  & $0^+(0^{++})$  &1000-1500&150$-$400 & 93.6  \\
$\rho(1450)$  & $1^+(1^{--})$  &1465   & 310    & seen \\ \hline \hline
\end{tabular}
\end{center}
\caption{Resonances included in the $\pi \pi \rightarrow \pi\pi$ channel as
listed in the PDG. Note that the $\sigma$ is not present in the
PDG and is not being described exactly as a {\it Breit-Wigner} shape;
we listed the fitted parameters shown in column 1 of Table 2 where
$G^\prime$ is the analog of the {\it Breit-Wigner} width.}
\end{table}
\noindent
the specific
ones which are included, together with their masses and widths, when
available from the Particle Data Group (PDG) \cite{pdg}  listings.

Essentially there are only three arbitrary parameters in the whole
model, these correspond to the three unknowns in the description of a
broad scalar resonance given by eq.~(\ref{sigma-propagator}) . We will
include only the minimal
two derivative chiral contact interaction contained in eq.~(\ref{Lag: sym}) of
Appendix~B.
Clearly, higher derivative contact interaction may also be included
(see, for example, sec. III.E of Ref.~\cite{Sannino-Schechter}).

As shown in Fig.~\ref{figura1}, although the introduction of the $\rho$
dramatically
improves unitarity up to about $2~GeV$, $R^0_0$ violates unitarity to
a lesser extent starting around $500~MeV$.
(As noted in Ref.~\cite{Sannino-Schechter}, the $I=J=0$ channel is the only
troublesome one.) To completely restore unitarity in the present framework it
is
necessary to include a low mass broad scalar state which has historically
been denoted as the $\sigma$. It seems helpful to recall the contribution
of such a particle to the real part of the amplitude component $A(s,t,u)$
defined in eq.~(\ref{eq:def}):

\begin{equation}
Re A_{\sigma}(s,t,u)=Re
\frac{32\pi}{3H}
\frac{G}{M^3_\sigma}(s-2m_\pi^2)^2
\frac{(M_\sigma^2-s)+i M_{\sigma}G^{\prime}}{(s-M_\sigma^2)^2 +
M_\sigma^2{G^\prime}^2}\ ,
\label{eq:sigma}
\end{equation}
where
\begin{equation}
H=\left(1-4\frac{m_{\pi}^2}{M^2_\sigma}\right)
 ^{\frac{1}{2}}
\left(1-2\frac{m_{\pi}^2}{M^2_\sigma}\right)^2\approx 1\ ,
\end{equation}
and $G$ is related to the coupling constant $\gamma_0$ defined in
eq.~(\ref{la:sigma}) by
\begin{equation}
G={\gamma}^2_0\frac{3 H M_{\sigma}^3}{64\pi}\ .
\label{sigma-coupling}
\end{equation}
Note that the factor $(s-2m_{\pi}^2)^2$ is due to the derivative-type coupling
required for chiral symmetry in eq.~(\ref{la:sigma}). The total amplitude will
be
crossing symmetric since $A(s,t,u)$ and $A(u,t,s)$ in eq.~(\ref{eq:def}) are
obtained
by performing the indicated permutations. $G^{\prime}$ is a parameter
which we introduce to regularize the propagator. It can be called a width,
but it turns out to be rather large so that, after the $\rho$ and $\pi$
contributions are taken into account, the partial wave amplitude
$R^0_0$ does not clearly display the characteristic resonant behavior. In the
most general situation one might imagine that $G$ could become complex
as in eq.~(\ref{rescattering}) due to higher order in $\displaystyle{1/N_c}$
corrections. It should be noted,
however, that eq.~(\ref{rescattering}) expresses nothing more than the
assumption of
unitarity for a {\it narrow} resonance and hence should not really be
applied to the present broad case. A reasonable fit was found in
Ref.~\cite{Sannino-Schechter} for $G$ purely real, but not equal to
$G^{\prime}$. By the use of eq.~(\ref{imaginary}), unitarity is in fact
locally satisfied.

A best overall fit is obtained with the parameter
choices; $M_{\sigma}=559~MeV$, $G/G^{\prime}=0.29$ and $G^{\prime}=370~MeV$ .
These have
been slightly fine-tuned from the values in Ref.~\cite{Sannino-Schechter}
in order to obtain a better fit in the $1~GeV$ region. The result for the
real part $R^0_0$ due to the inclusion of the $\sigma$ contribution along
with the $\pi$ and $\rho$ contributions is shown in Fig.~\ref{figura2}.
\figinsert{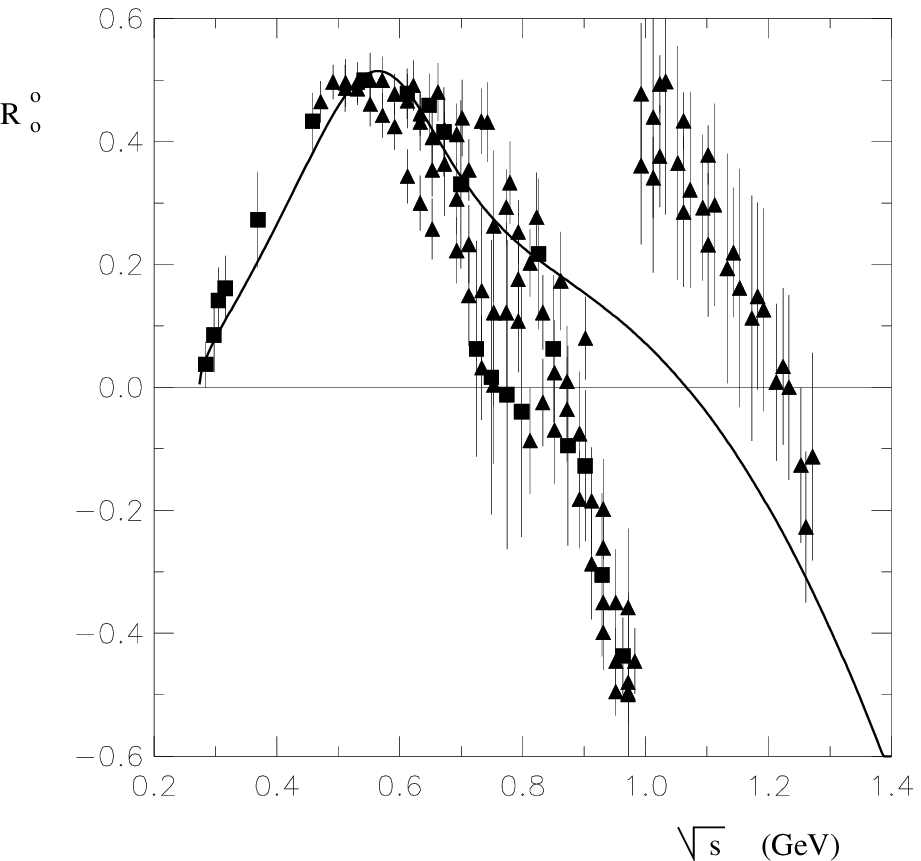}{The solid line is the {\it current algebra}
$+~\rho+\sigma$ result for $R^0_0$. The experimental points, in this
and suceeding figures, are
extracted from the phase shifts using eq.~(\ref{real-imaginary}) and
actually correspond to $R^0_0/\eta^0_0$. ($\Box$) are extracted from
the data of Ref.~\cite{lowenergy} while ($\triangle$) are extracted from the
data of Ref.~\cite{highenergy}. The predicted $R^0_0$ is small around the
$1~GeV$ region.}{figura2}{htpb}
It is seen
that the unitarity bound is satisfied and there is a reasonable agreement
with the experimental points \cite{lowenergy,highenergy} up to about $800~MeV$.
Beyond this point the effects of other resonances (mainly the $f_0(980)$)
are required. {}From eqs.~(\ref{eq:sigma}), (\ref{eq:isospin}) and
(\ref{eq:wave}) we see that the contribution of
$\sigma$ to $R^0_0$ turns
negative when $s>M^2_{\sigma}$. This is the mechanism which leads to
satisfaction of the unitarity bound (c.f. Fig.~\ref{figura1}). For
$s<M^2_{\sigma}$
one gets a positive contribution to $R^0_0$. This is helpful to push
the predicted curve upwards and closer to the experimental results
in this region, as shown in Fig.~\ref{figura3}.
\figinsert{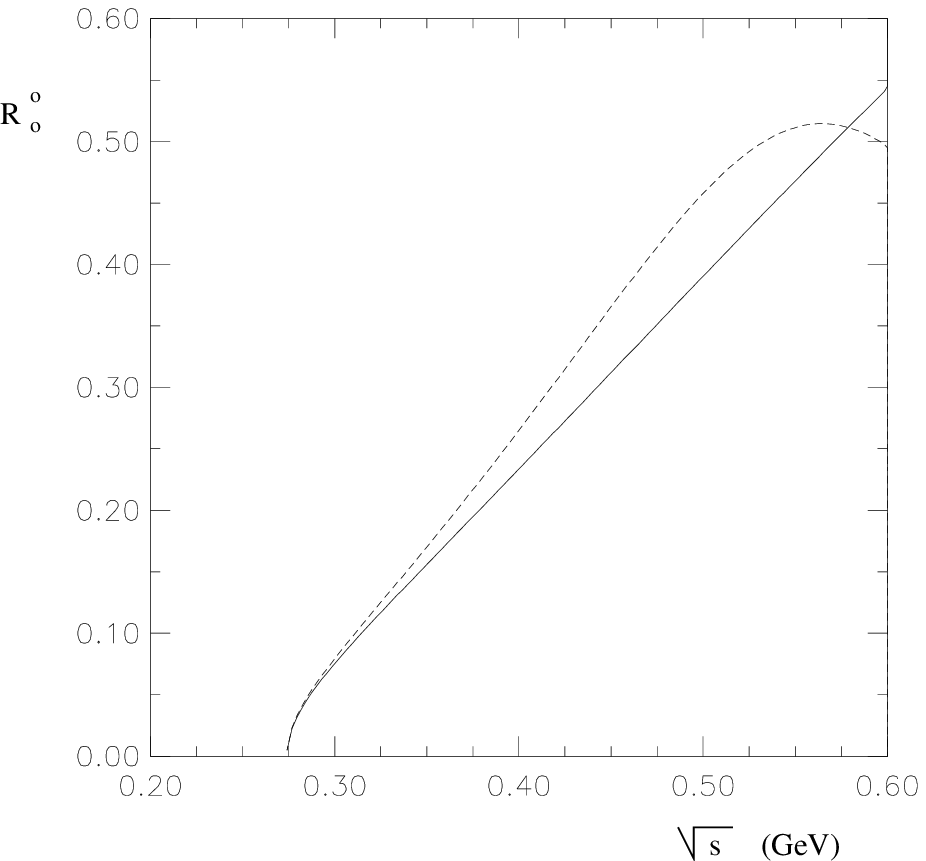}{A blowup of the low energy region. The solid line is
the {\it current algebra}
$+~\rho$ contribution to $R^0_0$. The dashed line includes the
$\sigma$ and has the effect of turning the curve down to avoid
unitarity violation while boosting it at lower energies.}{figura3}{htbp}
The four-derivative contribution in
the chiral perturbation theory approach performs the same function;
however it does not change sign and hence does not satisfy  the unitarity
bound above the $450~MeV$ region \cite{gasser-meissener}.

\section{The $1$ GeV Region}
\setcounter{equation}{0}

\subsection{The main point}

Reference to Fig.~\ref{figura2}  shows that the experimental data for
$R^0_0$ lie considerably lower than the $\pi+\rho+\sigma$ contribution between
$0.9$ and $1.0~GeV$ and then quickly reverse sign above this point.
We will now see that this distinctive shape is almost completely
explained by the inclusion of the relatively narrow scalar resonance $f_0(980)$
in a suitable manner. One can understand what is going on very simply
by starting from the real part of eq.~(\ref{rescattering}):
\begin{equation}
M\Gamma
\frac{(M^2-s)\cos (2\delta)-M\Gamma \sin (2\delta)}{(M^2-s)^2+M^2 {\Gamma}^2}
+ \frac{1}{2}\sin (2\delta)\ .
\label{980-mechanims}
\end{equation}
This expresses nothing more than the restriction of local unitarity in
the case of a narrow resonance
in the presence of a background. We have seen that the difficulty
of comparing the tree level $\displaystyle{1/N_c}$ amplitude to experiment is
enhanced
in the neighborhood of a direct channel pole. Hence it is probably
most reliable to identify the background term
$\displaystyle{\frac{1}{2} \sin(2\delta)}$ with our prediction for
$R^0_0$. In the region of interest, Fig.~\ref{figura2} shows that $R^0_0$ is
very
small so that one expects, $\delta$ to be roughly $90^\circ$ (assuming a
monotonically increasing phase shift). Hence
the first, pole term is approximately
\begin{equation}
-\frac{(M^2-s)M\Gamma}{(M^2-s)^2+M^2 {\Gamma}^2}\ ,
\label{980-pole}
\end{equation}
which contains a crucial reversal of sign compared to the real part of
eq.~(\ref{Breit-Wigner}). Thus, just below the resonance there is a sudden
{\it negative} contribution which jumps to a positive one above the
resonance. This is clearly exactly what is needed to bring experiment
and theory into agreement up till about $1.2~GeV$, as is shown
in Fig.~\ref{figura4}.
\figinsert{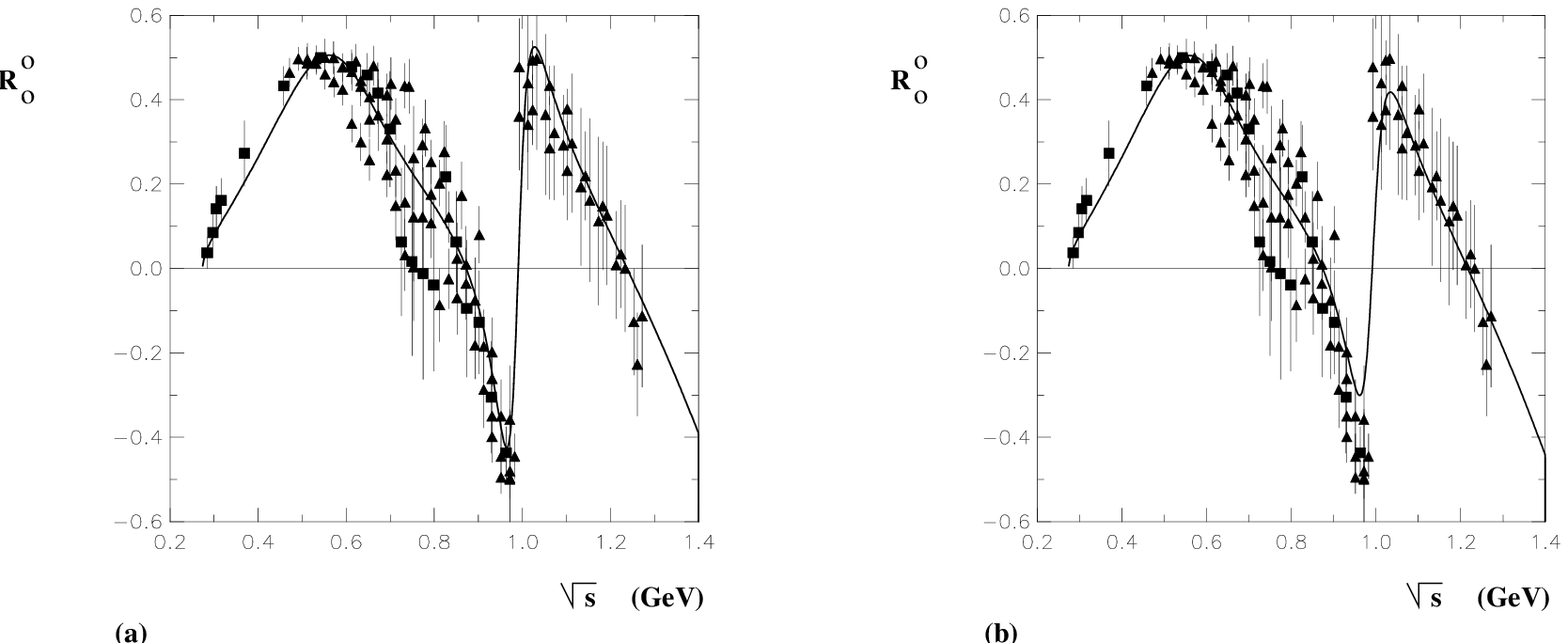}{(a): The solid line is the {\it current algebra}
$+~\rho~+~\sigma~+~f_0(980)$ result for $R^0_0$ obtained by assuming
column 1 in Table 2 for the $\sigma$ and $f_0(980)$ parameters
($Br(f_0(980)\rightarrow 2\pi)=100\%$).
(b): The solid line is the {\it current algebra}
$+~\rho~+~\sigma~+~f_0(980)$ result for $R^0_0$ obtained by assuming
column 2 in Table 2 ($Br(f_0(980)\rightarrow 2\pi)=78.1\%$) .}
{figura4}{htbp}
The actual amplitude used for this calculation properly
contains the effects of the pions' derivative coupling to the
$f_0(980)$ as in eq.~(\ref{eq:sigma}).

It is interesting to
contrast this picture with Fig.~10 in Ref.~\cite{Sannino-Schechter}.
There the interaction with the background was not taken into account
and there was no reversal of sign. Thus, although the unitarity bound
was obeyed, the experimental phase shifts could only be properly
predicted up to about $0.8~GeV$. If the $f_0(980)$ contribution in
that Fig.~10 is flipped in sign it is seen to agree with the present
Fig.~\ref{figura4}.

The above mechanism, which leads to a sharp dip in the $I=J=0$ partial
wave contribution to the $\pi\pi$-scattering cross section, can be
identified with the very old {\it Ramsauer-Townsend} effect
\cite{shiff} which concerned the scattering of $0.7~eV$ electrons on
rare gas atoms. The dip occurs because the background phase of $\pi/2$
causes the phase shift to go through $\pi$ (rather than $\pi/2$) at
the resonance position. (Of, course, the cross section is proportional
to $\sum_{I,J}^{~} (2J+1) \sin^2(\delta^J_I)$.) This simple mechanism seems
to be all that is required to understand the main feature of $\pi\pi$
scattering in the $1~GeV$ region.

\subsection{Detailed analysis}

Here we will compare with experimental data, the real part of the
$I=J=0$ partial wave amplitude which results from our crossing symmetric
model. First we will consider the sum of the contributions of the
{\it current algebra}, $\rho$-{\it meson}, $\sigma$ and $f_0(980)$
pieces. Then we will add pieces corresponding to the {\it next group}
of resonances; namely, the $f_2(1270)$, the $\rho (1450)$ and the
$f_0(1300)$. In this section we will continue to neglect the
$K\overline{K}$ channel.

The current algebra plus $\rho$ contribution to the quantity $A(s,t,u)$
defined in eq.~(\ref{eq:def}) is\footnote{We introduced the step function
$\theta(s-4 m_{\pi}^2)$ in the propagator and have checked that its
inclusion does not make much difference in the results.}
\begin{eqnarray}
A_{ca+\rho}(s,t,u) &=&2 \frac{s-m_{\pi}^2}{F_\pi ^2} +
\frac{g_{\rho\pi\pi}^2}{2m^2_{\rho}}(4 m_{\pi}^2 - 3 s)+\nonumber \\
&-& \frac{g_{\rho\pi\pi}^2}{2}
\left[\frac{u-s}{(m^2_{\rho}-t)-im_{\rho}{\Gamma}_{\rho}\theta (t - 4
m_{\pi}^2)} \right.\nonumber \\
&+& \left.
\frac{t-s}{(m^2_{\rho}-u)-im_{\rho}{\Gamma}_{\rho}\theta (u - 4
m_{\pi}^2)}\right]\ .
\label{ca+rho}
\end{eqnarray}
Note that for the $I=J=0$ channel this will yield a purely real
contribution to the partial wave amplitude. The contribution of the
low lying $\sigma$ meson was given in eq.~(\ref{eq:sigma}). For the
important $f_0(980)$ piece we have
\begin{equation}
Re A_{f_0(980)}(s,t,u)=Re\left[
\frac{\gamma_{f_0\pi\pi}^2 e^{2i\delta} (s-2 m_{\pi}^2)^2}
{m_{f_0}^2 - s - i m_{f_0}\Gamma_{tot}(f_0)\theta ( s - 4 m_{\pi}^2)}
\right]\ ,
\label{f0(980)}
\end{equation}
where $\delta$ is a background phase parameter and the real coupling
constant $\gamma_{f_0\pi\pi}$ is related to the $f_0(980)\rightarrow \pi\pi$
width by
\begin{equation}
\Gamma(f_0(980)\rightarrow
\pi\pi)=\frac{3}{32\pi}\frac{\gamma^2_{f_0\pi\pi}}
{m_{f_0}}\sqrt{1 - \frac{4 m_{\pi}^2}{m_{f_0}^2}}\ .
\end{equation}
We will not consider $\delta$ to be a new parameter but shall predict
it as
\begin{equation}
\frac{1}{2}\sin (2\delta)\equiv \tilde{R}^0_0 (s=m_{f_0}^2)\ ,
\end{equation}
where $\tilde{R}^0_0$ is computed as the sum of the current algebra,
$\rho$, and sigma pieces.
Since the $K\overline{K}$ channel is being neglected, one might want
to set the {\it regularization parameter}  $\Gamma_{tot}(f_0)$ in the
denominator to $\Gamma(f_0(980)\rightarrow \pi\pi)$. We shall try both this
possibility as well as the experimental one
$\displaystyle{\frac{\Gamma(f_0(980)\rightarrow
\pi\pi)}{ \Gamma_{tot}(f_0)}\approx 78.1\%}$.

A best fit of our parameters to the experimental data results in the
curves shown in Fig.~\ref{figura4}
for both choices of branching ratio. Only the
three parameters $G/G^{\prime}$, $G^{\prime}$ and $M_{\sigma}$ are
essentially free. The others are restricted by experiment.
Unfortunately the total width $\Gamma_{tot}(f_0)$ has a large
uncertainty; it is claimed by the PDG to lie in the $40-400~MeV$
range. Hence this is effectively a new parameter. In addition we have
considered the precise value of $m_{f_0}$ to be a parameter for fitting
purposes. The parameter values
for each fit are given in Table 2 together with the $\chi^2$ values.
It is clear that the fits are good and that the parameters are stable
against variation of the branching ratio. The predicted background
phase is seen to be close to $90^\circ$ in both cases. Note that the
fitted width of the $f_0(980)$ is near the low end of the experimental
range. The low lying sigma has a mass of around $560~MeV$ and a width
of about $370~MeV$. As explained in section 3, we are not using
exactly a conventional {\it Breit-Wigner} type form for this very
broad resonance. The numbers characterizing it do however seem reasonably
consistent with other determinations
\cite{Tornqvist:95,Janssen-Pearce-Holinde-Speth,Morgan-Pennington:93}.

\begin{table}

\begin{center}
\begin{tabular}{|c||c c c c||c c c c||c|} \hline \hline
\multicolumn{1}{|c||}{}&\multicolumn{4}{|c||}{}&
\multicolumn{4}{|c||}{With Next Group}&
\multicolumn{1}{|c|}{ No $\rho(1450)$}\\

{\small $BR(f_0(980)\rightarrow 2\pi)\%$}            & 100   & 78.1  & 78.1 &
78.1  &
100 & 78.1 & 78.1 & 78.1 & 100 \\

$\eta^0_0 $         & 1     & 1   & 0.8     & 0.6  & 1   & 1    & 0.8
& 0.6 & 1
\\ \hline \hline

$M_{f_0 (980)}~(MeV)$ & 987 & 989 & 990& 993  & 991 &992 &993 & 998 &992 \\

$\Gamma_{tot}~(MeV)$  &64.6  &  77.1& 75.9& 76.8  &66.7 &77.2 &78.0&84.0&64.6
\\

$M_{\sigma}~(MeV) $   &559  & 557 & 557& 556  &537& 537 & 535 & 533 & 525 \\

$G^\prime ~(MeV)$   &370  & 371 & 380& 395  &422& 412 & 426 & 451 &467 \\

$G/G^\prime$   & 0.290  & 0.294 & 0.294& 0.294  &0.270& 0.277
&0.275&0.270 & 0.263\\ \hline \hline

$\delta $~(deg.)      & 85.2  & 86.4   & 87.6 & 89.6 &   89.2  & 89.7  &
91.3&94.4  &90.4 \\

$\chi^2 $      & 2.0   & 2.8   &  2.7  & 3.1 &   2.4  & 3.2   & 3.2 &
3.4 & 2.5\\ \hline \hline
\end{tabular}

\end{center}
\caption{Fitted parameters for different cases of interest.}
\end{table}

\subsection{Effect of the next group of resonances}

Going up in energy we encounter $J^{PC}=2^{++}, 0^{++}$ and $1^{--}$
resonances in the $1300~MeV$ region. The properties of the $2^{++}$
state $f_2(1270)$ are very well established. For the others there is
more uncertainty but the PDG lists the $f_0 (1300)$ and $\rho (1450)$
as established states. However the mass of the $f_0 (1300)$ can
apparently lie anywhere in the $1000-1500~MeV$ range. In
Ref.~\cite{Sannino-Schechter} it was noted that the contributions of these
{\it next group} particles tended to cancel among themselves. Thus we
do not expect their inclusion to significantly change the previous
results in the range of interest up to about $1.2~GeV$.

In Fig.~\ref{figura5}
\figinsert{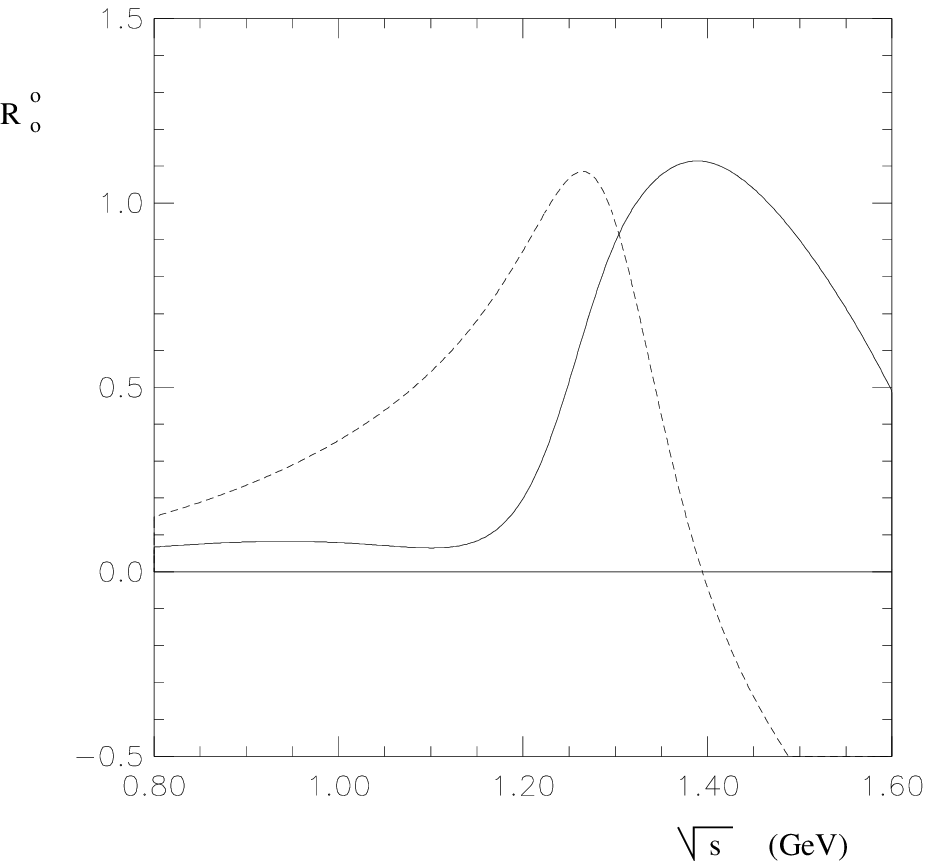}{Contribution from the {\it next group} of
resonances; the solid curve is obtained with the reverse sign of the
$f_0(1300)$ piece.}{figura5}{htbp}
we display the contribution of the {\it next group}
particles by themselves to $R^0_0$. (The amplitudes are summarized in
Appendix~C). The dashed curve is essentially a reproduction of Fig.~6
of Ref.~\cite{Sannino-Schechter}.
The somewhat positive net contribution of these resonances to $R^0_0$
is compensated by readjustment of the parameters describing the low
lying sigma.
It may be interesting to include the
effect of the background phase for the $f_0 (1300)$ as we have just
seen that it was very important for the proper understanding of the
$f_0 (980)$. To test this possibility we reversed the sign of the
$f_0 (1300)$ contribution and show the result as the solid curve in
Fig.~\ref{figura5}. This sign reversal is reasonable since our model suggests a
background phase of about $270^\circ$ in the vicinity of the $f_0
(1300)$. It can be seen that there is now a significantly greater
cancellation of the {\it next group} particles among themselves up to
about $1.2~GeV$. The resulting total fits are shown in Fig.~\ref{figura6}
\figinsert{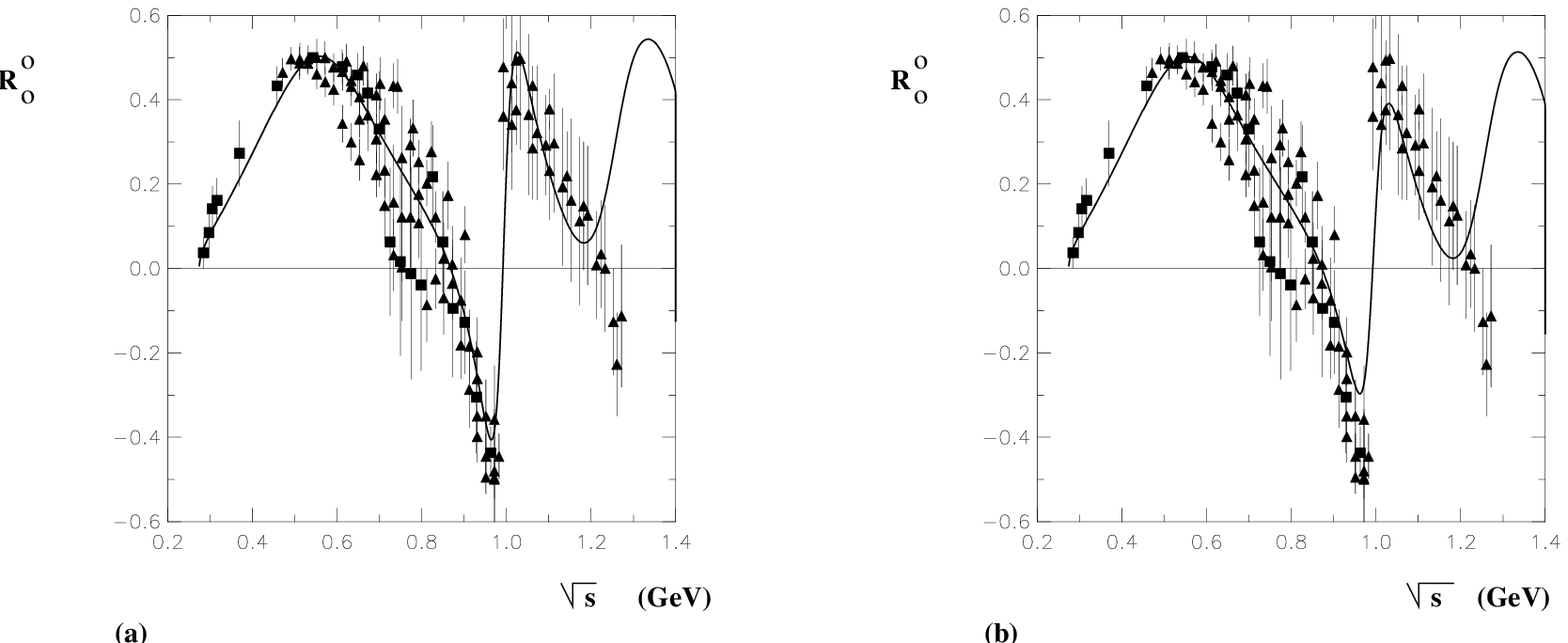}{Prediction for $R^0_0$ with the {\it next
group} of resonances. (a) assumes (column 5 in Table 2)
($BR(f_0(980)\rightarrow 2\pi)=100\%$) while (b) assumes (column 6)
($BR(f_0(980)\rightarrow 2\pi)=78.1\%$).}{figura6}{htpb}
 for both
$100\%$ and $78.1\%$ assumed $f_0 (980)\rightarrow \pi\pi$ branching ratios
and the parameters associated with the fits are shown in Table 2. It
is clear that the fitted parameters and results up to about $1.2~GeV$
are very similar to the cases when the {\it next group} was absent.
Above this region, there is now, however, a positive bump in $R^0_0$
at around $1.3~GeV$. This could be pushed further up by choosing a
higher mass (within the allowable experimental range) for the $f_0
(1300)$. Resonances in the $1500~MeV$ region, which have {\it not}
been taken into account here, would presumably also have an important
effect in the region above $ 1.2~GeV$. Clearly, there is not much
sense, at the present stage, in trying to produce a fit above
$1.2~GeV$.

The analysis above assumed that the $\rho(1450)$ decays predominantly
into two pions since the PDG listing does not give any specific
numbers. On the other hand the $K^*(1410)$, which presumably is in the
same $SU(3)$ multiplet as the $\rho(1450)$, has only a $7\%$ branching
ratio into $K\pi$. Thus it is possible that $\rho(1450)$ actually has
a small coupling to $\pi\pi$. To test this out we redid the
calculation with the complete neglect of the $\rho(1450)$
contribution. The resulting fit is shown in the last column of Table~2
and it is seen to leave the other parameters essentially unchanged.

It thus seems that the results are consistent with the
hypothesis of {\it local cancellation}, wherein the physics up to a
certain energy $E$ is described by including only those resonances up
to slightly more than $E$ and it is furthermore hypothesized that the
individual
particles cancel in such a way that unitarity is maintained.

\subsection{Effects of inelasticity}

Up to now we have completely neglected the effects of coupled
inelastic channels. Of course the $4\pi$ channel opens at $540~MeV$,
the $6\pi$ channel opens at $810~MeV$ and, probably most
significantly, the $K\overline{K}$ channel opens at $990~MeV$. We have
seen that a nice undestanding of the $\pi\pi$ elastic channel up to
about $1.2~GeV$ can be gotten with complete disregard of inelastic
effects. Nevertheless it is interesting to see how our results would
change if experimental data on the elasticity parameter $\eta^0_0$ are
folded into the analysis. Figure~\ref{figura7}
\figinsert{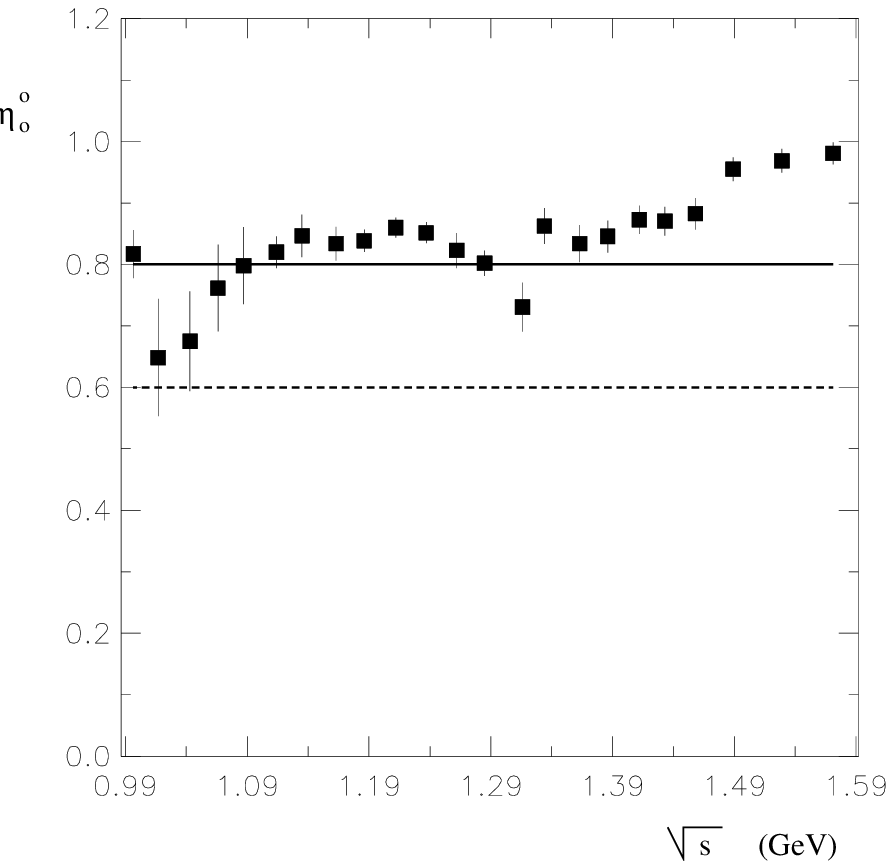}{An experimental determination of $\eta^0_0=
\sqrt{1 - 4 |T^0_{12,0}|^2}$ \cite{Cohen:80}.}{figura7}{htpb}
illustrates the results for
$\eta^0_0(s)$ obtained from an experimental analysis \cite{Cohen:80}
of $\pi\pi\rightarrow K\overline{K}$ scattering. For simplicity, we
approximated the data by a constant value $\eta^0_0 = 0.8$ above the
$K\overline{K}$ threshold. Figure~\ref{figura8}(a)
\figinsert{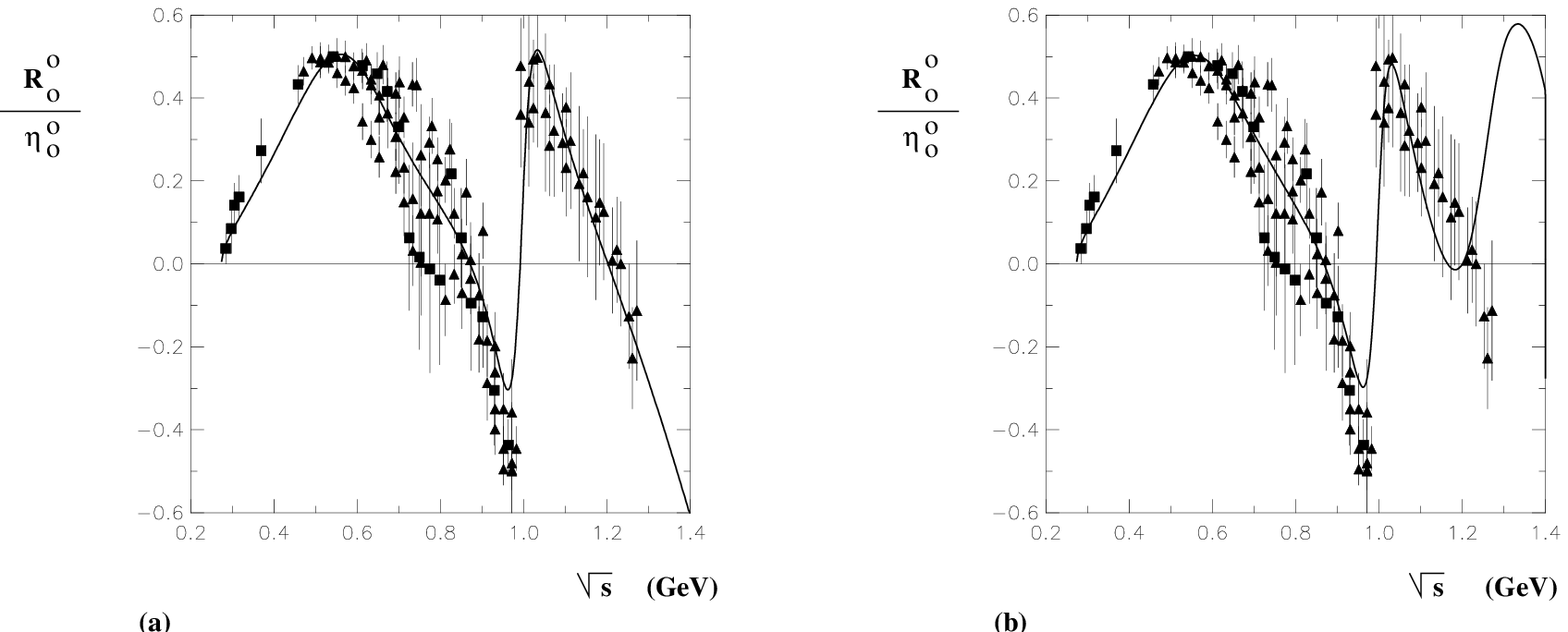}{Predictions with phenomenological treatment of
inelasticity ($\eta^0_0 = 0.8$) above $K\overline{K}$ threshold. (a):
without {\it next group}. (b): with {\it next group}. }{figura8}{htbp}
shows the effect of this choice on
$R^0_0(s)$ computed without the inclusion of the {\it next group} of
resonances, while Fig.~\ref{figura8}(b) shows the effect when the {\it next
group}
is included. Comparing with Fig.~\ref{figura4}(b) and \ref{figura6}(b), we see
that setting $\eta^0_0 =
0.8$ has not made any substantial change. The parameters of the fit
are shown in Table 2
as are the parameters for an alternative fit with
$\eta^0_0 = 0.6$. The latter choice leads to a worse fit for $R^0_0$.

We conclude that inelastic effects are not very important for
understanding the main features of $\pi\pi$ scattering up to about
$1.2~GeV$. However, we will discuss the calculation of $\eta^0_0 (s) $
from our model in section 5.

\subsection{Phase shift}

Strictly speaking our initial assumption only entitles us
to compare, as we have already
done, the real part of the predicted amplitude with the real part of
the amplitude deduced from experiment. Since the predicted $R^0_0(s)$
up to $1.2~GeV$ satisfies the unitarity bound (within the fitting
error) we can calculate the imaginary part $I^0_0(s)$, and hence the
phase shift $\delta_0^0(s)$ on the assumption that full unitarity
holds. This is implemented by substituting $R^0_0(s)$ into
eq.~(\ref{imaginary}) and resolving the discrete sign ambiguities by
demanding that $\delta^0_0(s)$ be continuous and monotonically
increasing (to agree with experiment). It is also necessary to know
$\eta^0_0(s)$ for this purpose; we will be content with the
approximations above which seem sufficient for understanding the main
features of $\pi\pi$ scattering up to $1.2~GeV$.

In this procedure there is a practical subtlety already discussed at the
end of section IV of Ref.~\cite{Sannino-Schechter}. In order for
$\delta^0_0(s)$ to increase monotonically it is necessary that the
sign in front of the square root in eq.~(\ref{imaginary}) change. This can
lead to a discontinuity unless $2|R^0_0(s)|$ precisely reaches
$\eta^0_0(s)$. However the phase shift is rather sensitive to small
deviations from this exact matching. Since the fitting procedure does
not enforce that $|R^0_0(s)|$ go precisely to $\eta^0_0(s)/2 \approx
0.5$, this results in some small discontinuities. (These could be
avoided by trying to fit the phase shift directly.)

Figure~\ref{figura9}
\figinsert{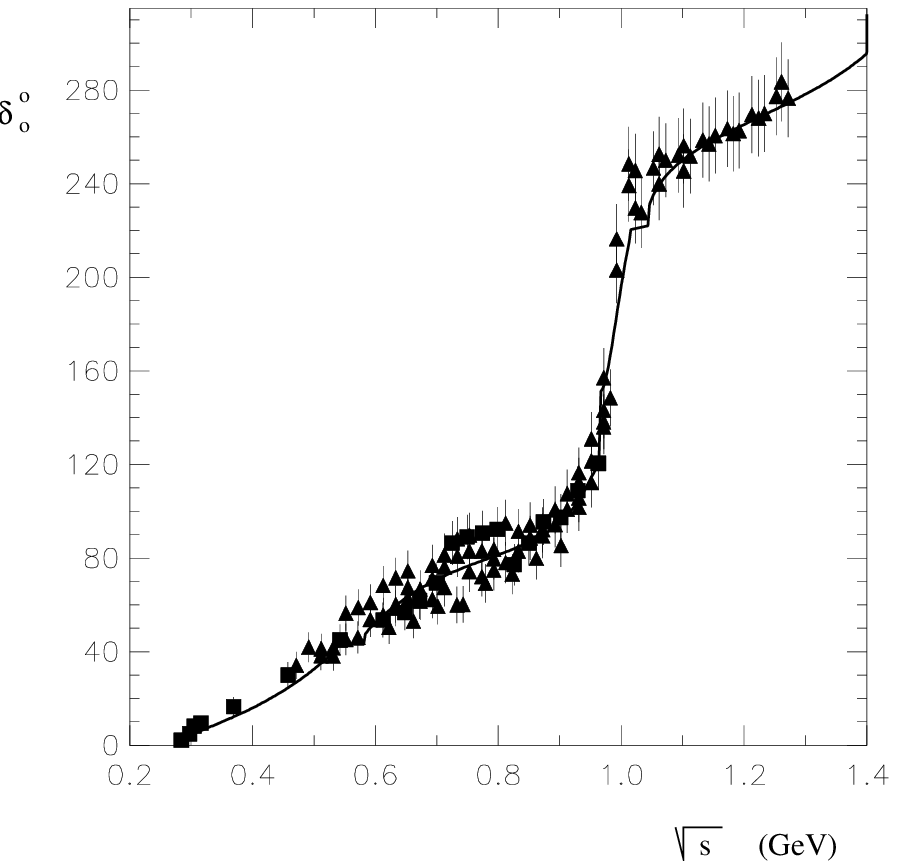}{Estimated phase shift using the
predicted real part and unitarity relation.}{figura9}{htpb}
shows the phase shift $\delta^0_0(s)$ estimated in this manner
for parameters in the first column of Table 2. As expected, the
agreement is reasonable. A very similar estimate is obtained when
(column 3 of Table 2) $\eta^0_0$ is taken to be $0.8$ while considering
the $\pi\pi$ branching ratio of $f_0(980)$ to be its experimental
value of $78.1\%$. It appears that these two parameter changes are
compensating each other so that one may again conclude that the
turning on of the $K\overline{K}$ channel really does not have a major
effect. When the {\it next group} of resonances is included (column 7
of Table 2) the estimated $\delta^0_0(s)$ is very similar up to about
$1.2~GeV$. Beyond this point it is actually somewhat worse, as we
would expect by comparing Fig.~\ref{figura8}(b) with Fig.~\ref{figura8}(a).

\section{$\pi\pi \rightarrow K\overline{K}$ Channel}
\setcounter{equation}{0}

We have seen that $\pi\pi \rightarrow \pi \pi $ scattering can be understood
up to about $1.2~GeV$ with the neglect of this inelastic channel.  In
particular, a phenomenological description of the inelasticity did not
change the overall picture. However we would like to begin to explore
the predictions of the present model for this channel also. The whole
coupled channel problem is a very complicated one so we will be
satisfied here to check that the procedure followed for the $\pi\pi$
elastic channel can lead to an inelastic amplitude which also
satisfies the unitarity bounds. Specifically we will confine our
attention to the real part of the $I=J=0$  $\pi\pi \rightarrow
K\overline{K}$ amplitude, $R^0_{12;0}$ defined in eq.~(\ref{eq:wave}).

In exact analogy to the $\pi\pi \rightarrow \pi\pi$ case we first consider
the contribution of the contact plus the $K^*(892)$ plus the
$\sigma(550)$ terms. It is necessary to know the coupling strength of
the $\sigma$ to $K\overline{K}$, defined by the effective Lagrangian
piece
\begin{equation}
-\frac{\gamma_{\sigma K \overline{K}}}{2}\sigma \partial_\mu
{\overline{K}}
\partial_\mu {K}\ .
\end{equation}
If the $\sigma$ is ideally mixed and there is no OZI rule violating
piece we would have $\gamma_{\sigma K \overline{K}}=\gamma_0$ as
defined in eq.~(\ref{la:sigma}). For definiteness, we shall adopt this
standard mixing assumption. The appropriate amplitudes are listed in
Appendix~C. Figure~\ref{figura10}
\figinsert{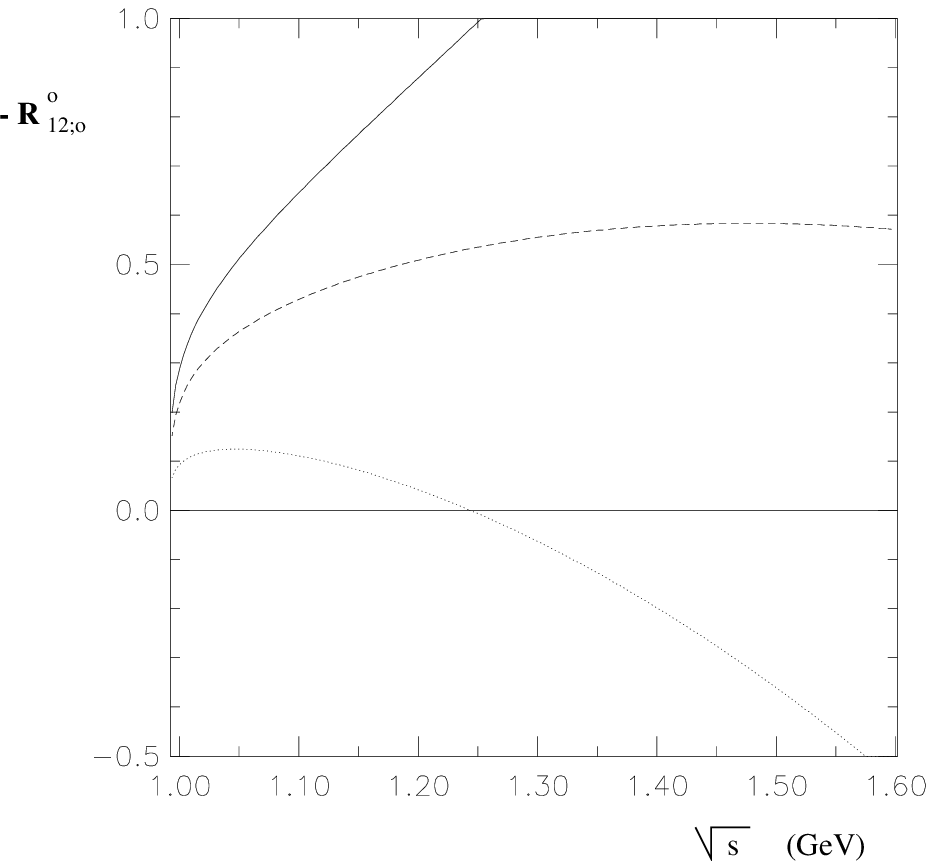}{Contributions to $\pi\pi \rightarrow K\overline{K}$
($R^0_{12;0}$). The solid line shows the current algebra result; the
dashed line represents the inclusion of $K^*(892)$; the dotted line
includes the $\sigma(550)$ too.}{figura10}{hptb}
shows the plots of  $R^0_{12;0}$ for the
current algebra part alone, the current algebra plus $K^*$ and the
current algebra plus $K^*$ plus $\sigma$ parts. Notice that
unitarity requires
\begin{equation}
| R^0_{12;0}|\leq\frac{\sqrt{1-{\eta^0_0}^2}}{2}\leq \frac{1}{2}\ .
\end{equation}
The current algebra result already clearly violates this bound at
$1.05~GeV$. As before, this is improved by the $K^*$ vector meson
exchange contribution and further improved by the very important tail
of the $\sigma$ contribution. The sum of all three shows a structure
similar to the corresponding Fig.~\ref{figura2} in the $\pi\pi \rightarrow
\pi\pi$ case.
The unitarity bound is not violated until about $1.55~GeV$.

Next, let us consider the contribution of the $f_0(980)$ which, since
the resonance straddles the threshold, is expected to be important. We
need to know the effective coupling constant of the $f_0$ to $\pi\pi$
and to $K\overline{K}$. As we saw in eq.~(\ref{f0(980)}), and the
subsequent discussion, the effective $\pi\pi$ coupling should be taken
as $\gamma_{f_0\pi\pi} e^{i\frac{\pi}{2}}$. Experimentally, only the
branching ratios for $f_0(980)\rightarrow \pi\pi$ and $f_0(980)\rightarrow
K\overline{K}$ are accurately known. We will adopt for definiteness
the value of $\gamma_{f_0\pi\pi}$ corresponding to the fit in the third
column of Table 2 ($\Gamma_{tot}(f_0(980))=76~MeV$). It is more
difficult to estimate the $f_0(980)\rightarrow K\overline{K}$ effective
coupling constant since the central value of the resonance may
actually lie {\it below} the threshold. By taking account
\footnote{With $\Gamma_{tot}(f_0(980))=76~MeV$ we would have $\Gamma
(f_0(980)\rightarrow K\overline{K}))=16.6~MeV$. Then $\gamma_{f_0 K
\overline{K}}$ is estimated from the formula:
\begin{displaymath}
16.6~MeV=|\gamma_{f_0 K \overline{K}}|^2\int_{2m_k}^{\infty}
\rho(M) |A(f_0(M)\rightarrow K\overline{K})|^2 \Phi (M)\,dM\ ,
\end{displaymath}
where $A(f_0(M)\rightarrow K\overline{K})$ is the reduced amplitude for an
$f_0$ of mass M to decay to $K\overline{K}$, $\Phi(M)$ is the phase
space factor and $\rho(M)$ is the weighting function given by
\begin{displaymath}
\rho(M)=\sqrt{\frac{2}{\pi}}\frac{1}{\Gamma_{tot}}exp\left\{-2\left[\frac
{(M-M_0)^2}{\Gamma_{tot}^2}\right] \right\}\ .
\end{displaymath}
Here, $M_0$ is the central mass value of the $f_0(980)$.}
of the finite
width of the $f_0(980)$ we get the rough estimate
$|\gamma_{f_0 K \overline{K}}|=10~GeV^{-1}\approx 4
|\gamma_{f_0\pi\pi}|$ for the choice in the third column,
$M_{f_0(980)}=990~MeV$. Of course, this estimate is very sensitive to
the exact value used for $M_{f_0(980)}$. It seems reasonable to take
$\gamma_{f_0 K \overline{K}}$ to be purely real. The results of
including the $f_0(980)$ contribution, for both sign choices of $
 \gamma_{f_0 K \overline{K}}$, are shown in Fig.~\ref{figura11}.
\figinsert{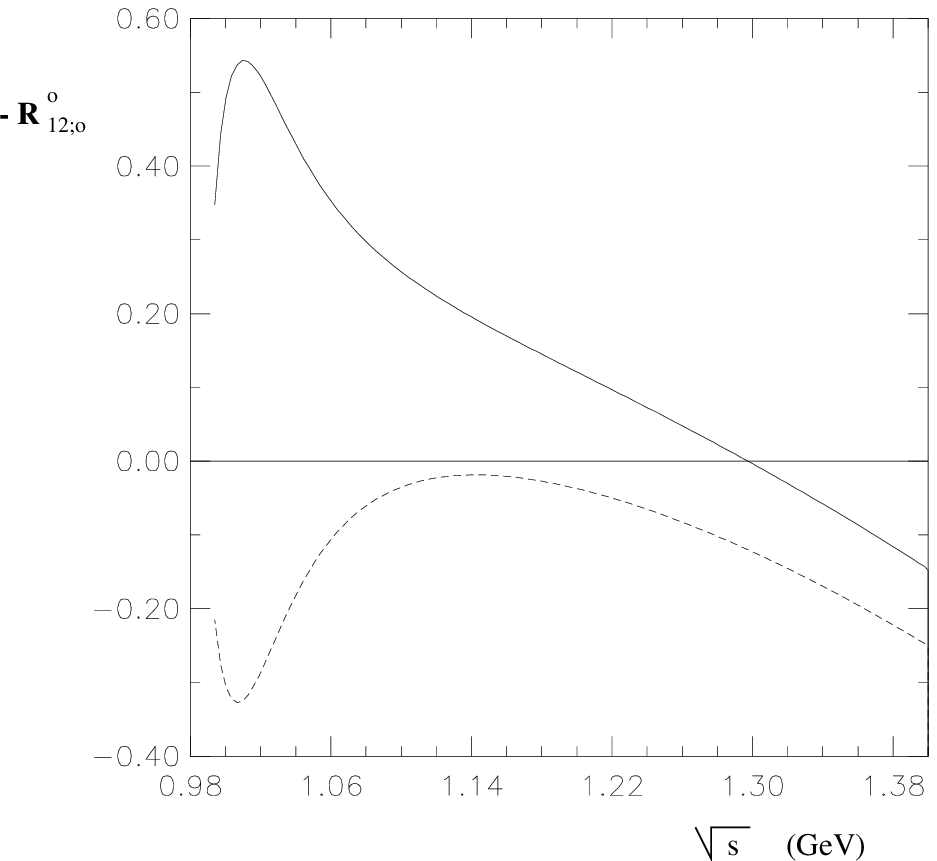}{Effect of $f_0(980)$ on $\pi\pi \rightarrow
K\overline{K}$. The solid curve corresponds to a negative $\gamma_{f_0
K \overline{K}}$ and the dashed one to a positive sign.}{figura11}{htpb}
The unitarity
bounds are satisfied for the positive sign of $\gamma_{f_0 K
\overline{K}}$ but slightly violated for the negative sign choice.

Finally, let us consider the contributions to $\pi\pi \rightarrow
K\overline{K}$ from the members of the multiplets containing the {\it
next group} of particles. There will be a crossed channel contribution
from the strange excited vector meson $K^*(1410)$.  However it will be
very small since $K^*(1410)$ predominately couples to $K^*\pi$ and
has only a $7\%$ branching ratio to $K\pi$. In addition there will be
a crossed channel scalar $K^*_0(1430)$ diagram as well as a direct
channel scalar $f_0(1300)$ diagram contributing to $\pi\pi \rightarrow
K\overline{K}$. The $f_0(1300)$ piece is small because $f_0(1300)$ has
a very small branching ratio to $K\overline{K}$. Furthermore the
$K^*_0(1430)$ piece turns out also to be small; we have seen that the
crossed channel scalar gave a negligible contribution to $\pi\pi \rightarrow
\pi\pi$. The dominant {\it next group} diagrams involve the tensor
mesons. Near threshold, the crossed channel $K^*_2(1430)$ diagram is
the essential one since the direct channel $f_2(1270)$ contribution
for the $J=0$ partial wave is suppressed by a spin-2 projection
operator. Above $1270~MeV$ the $f_2(1270)$ contribution becomes
increasingly important although it has the opposite sign to the
crossed channel tensor piece.
Figure~\ref{figura12} shows the net prediction for $R_{12;0}^0$ obtained with
the inclusion of the main {\it next
group} contributions from the $K^*_2(1430)$ and $f_2(1270)$.
\figinsert{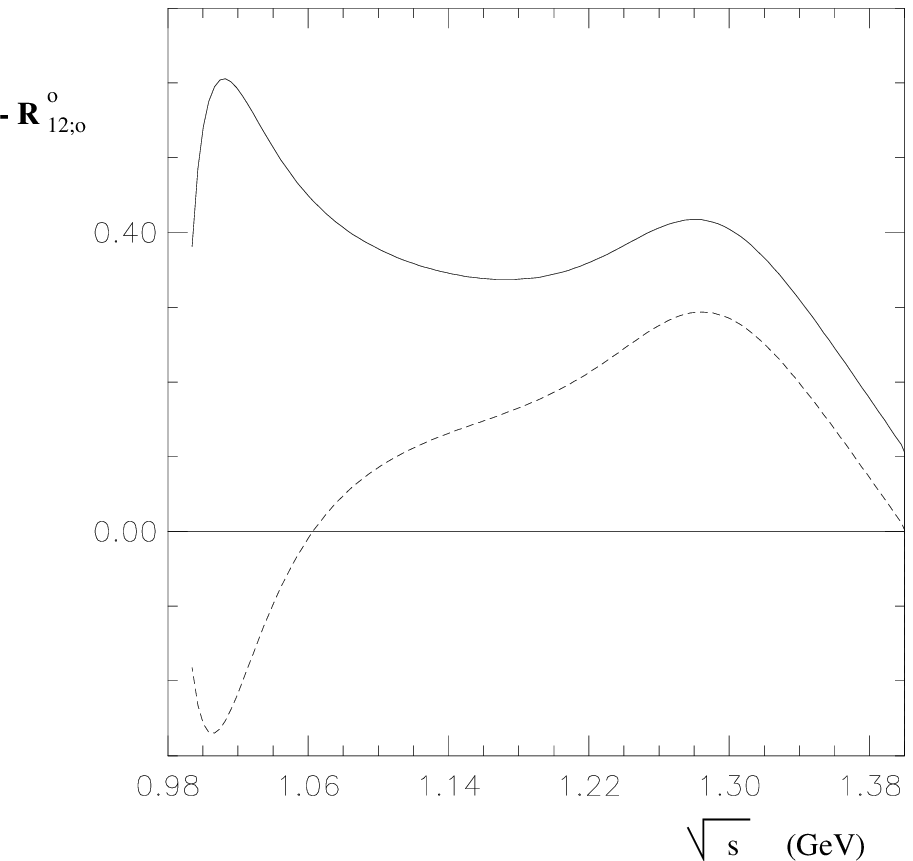}{Effects on $\pi\pi \rightarrow K \overline{K}$ due
to the {\it next group} of resonances for the two different sign
choices in Fig. \ref{figura11}.}{figura12}{htpb}
Both
assumed signs for $\gamma_{f_{0} K\overline{K}}$ are shown and other
parameters correspond to column 3 of Table 2. Clearly there is an
appreciable effect. Figure~\ref{figura13}
\figinsert{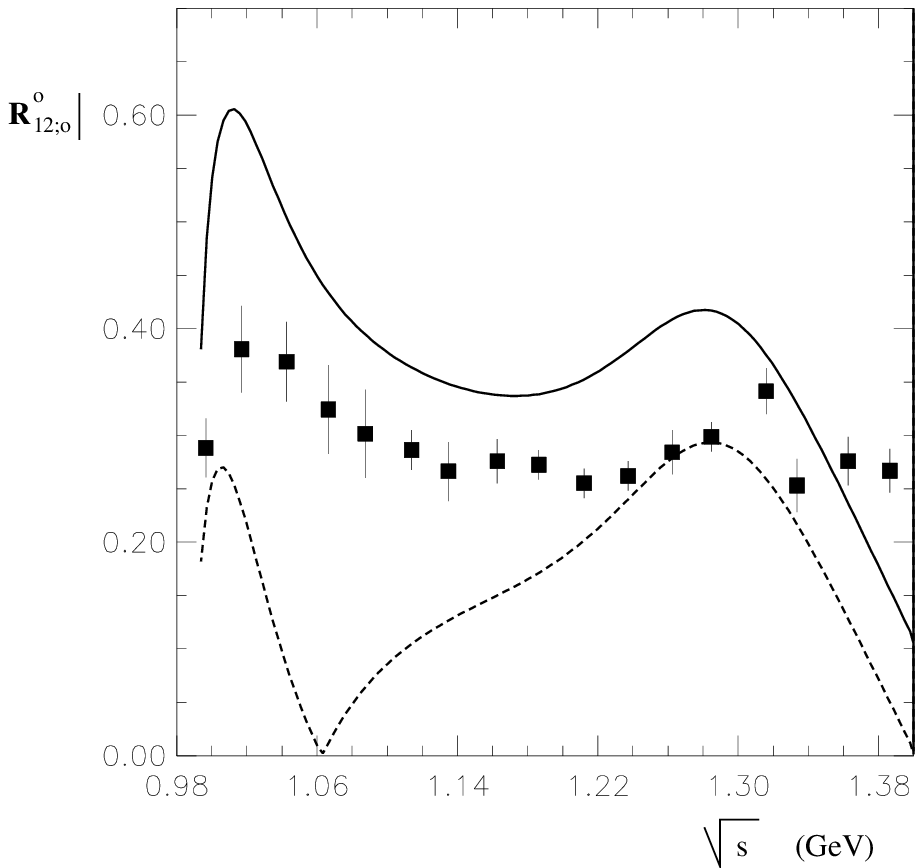}{ $|R_{12;0}^0|$ together with one experimental
determination \cite{Cohen:80} of
$\displaystyle{|T_{12;0}^0|}=\sqrt{(R_{12;0}^0)^2+(I_{12;0}^0)^2}$.
Signs for $\gamma_{f_0 K \overline{K}}$ as in
Fig.~\ref{figura11}}{figura13}{htpb}
shows the magnitude of $|R_{12;0}^0|$ together with one experimental
determination \cite{Cohen:80} of
$\displaystyle{|T_{12;0}^0|}=\sqrt{(R_{12;0}^0)^2+(I_{12;0}^0)^2}$. The
positive sign of  $\gamma_{f_0K\overline{K}}$ is favored but,
considering the uncertainty in  $|\gamma_{f_0K\overline{K}}|$ among
other things, we shall not insist on this. It seems to us that the
main conclusion is that the unitarity bound can be satisfied in the
energy range of interest.

\section{Summary and Discussion}
\setcounter{equation}{0}

We have obtained a simple approximate analytic form for the real part
of the $\pi\pi$
scattering amplitude in the energy range from threshold to about
$1.2~GeV$. It satisfies both crossing symmetry and (more
non-trivially) unitarity in this range. Inspired by the leading
$\displaystyle{1/N_c}$
approximation, we have written the amplitude as the sum of a contact
term and poles. Of course the leading $\displaystyle{1/N_c}$ amplitude can not
be
directly compared with experiment since it is purely real (away from
the direct channel poles) and diverges at the pole positions.
Furthermore, an infinite number of poles, and higher derivative
interactions are in principle needed. To overcome these problems we
have employed the following procedure.
\begin{itemize}
\item[a.]{We specialized to predicting the real part of the amplitude.}
\item[b.]{We postulated that including only resonances from threshold
to slightly more than the maximum energy of interest is sufficient. We
have seen that this {\it local cancellation} appears stable under the
addition of resonances in the $1300~MeV$ range. Beyond this range we
would expect still higher resonances to add in such a way so as to
enforce unitarity at still higher energies. }
\item[c.]{In the effective interaction Lagrangian we included only
terms with the minimal number of derivatives consistent with the
assumed chiral symmetry.}
\item[d.]{The most subtle aspect concerns the method for regularizing
the divergences at the direct channel resonance poles. In the simplest
case of a single resonance dominating a particular channel (e.g. the
$\rho$ meson) it is sufficient to add the standard {\it width} term to
the denominator (e.g. the real part of eq.~(\ref{Breit-Wigner})). For an
extremely broad resonance (like a needed low energy scalar isosinglet)
the concept of {\it width} is not so clear and we employed the slight
modification of the Breit-Wigner amplitude given in
eq.~(\ref{sigma-propagator}). Finally, for a relatively narrow resonance
in the presence of a non-negligible background we employed the
regularization given in eq.~(\ref{rescattering}) which includes the
background phase. Self-consistency is assured by requiring that the
background phase should be predicted by the model itself.}
\end{itemize}

All the regularizations introduced above are formally of higher than
leading order in the $\displaystyle{1/N_c}$ expansion (i.e. of order $1/N_c^2$
and
higher) and correspond physically to rescattering effects. In the case
of non-negligible background phase, there is an interesting difference
from the usual tree-level treatment of pole diagrams. The effective
squared coupling constant, $g^2_{R\pi\pi}$ of such a resonance to two
pions, is then not necessarily real positive. Since this
regularization is interpreted as a rescattering effect it does not
mean that ghost fields are present in the theory. This formulation
maintains crossing symmetry which is typically lost when a
unitarization method is employed.

In this analysis, the most non-trivial point is the satisfaction of
the unitarity bound for the predicted real part of the partial wave
amlitude,
\begin{equation}
|R^I_l| \leq \frac{\eta^I_l}{2}\ ,
\end{equation}
where $\eta^I_l < 1 $ is the elasticity parameter. The well known
difficulty concerns $R^0_0$. If $\eta^I_l (s)$ is known or calculated
the imaginary part $I^I_l (s)$ can be obtained, up to discrete
ambiguities, by eq.~(\ref{imaginary}).

The picture of $\pi\pi$ scattering in the threshold to slightly more
than $1~GeV$ range which emerges from this model has four parts. Very
near threshold the current algebra contact term approximates
$R^0_0(s)$ very well. The imaginary part $I^0_0(s)$, which is formally
of order $1/N^2_c$ can be obtained from unitarity directly using
eq.~(\ref{imaginary}) or, equivalently, by chiral perturbtion theory.
At somewhat higher energies the most prominent feature is the $\rho$
meson pole in the $I=J=1$ channel. The crossed channel $\rho$ exchange
is also extremely important in taming the elastic unitarity violation
associated with the current algebra contact term (Fig.~\ref{figura1}). Even
with
the $\rho$ present, Fig.~\ref{figura1} shows that unitarity is still violated,
though much less drastically. This problem is overcome by introducing
a low mass $\approx 550~MeV$, extremely broad sigma meson. It also has
another desirable feature: $R^0_0(s)$ is boosted
(see Fig.~\ref{figura3}) closer
to experiment in the $400-500~MeV$ range. The three parameters
characterizing this particle are essentially the only unknowns in the
model and were determined by making a best fit. In the $1~GeV$ region
it seems clear that the $f_0(980)$ resonance, interacting with the
predicted background in the manner of the {\it Ramsauer-Townsend}
effect, dominates the structure of the $I=J=0$ phase shift. The
inelasticity associated with the opening of the $K\overline{K}$
threshold has a relatively small effect. However we also presented a
preliminary calculation which shows that the present approach
satisfies the unitarity bounds in the inelastic $\pi\pi \rightarrow K
\overline{K}$ channel.

Other recent works
\cite{Tornqvist:95,Janssen-Pearce-Holinde-Speth,Morgan-Pennington:93,Harada,lnc} which approach
the problem in different ways, also contain a low mass broad sigma. The
question of whether the lighter scalar mesons are of $q\overline{q}$ type or
{\it meson-meson} type has also been discussed
\cite{Tornqvist:95,Janssen-Pearce-Holinde-Speth,Morgan-Pennington:93}.
In our model it is difficult to decide this issue. Of course, it is
not a clean question from a field theoretic standpoint. This question
is important for understanding whether the contributions
of such resonances are formally leading in the $\displaystyle{1/N_c}$
expansion. We are
postponing the answer as well as the answer to how to derive the
rescattering effects that were used to {\it regularize} the amplitude
near the direct channel poles as higher order in $\displaystyle{1/N_c}$
corrections.
Presumably, the rescattering effects could some day be calculated as
loop corrections with a (very complicated) effective Wilsonian action.
This would be a generalization of the chiral perturbation scheme of pions.
Another aspect of the $\displaystyle{1/N_c}$ picture concerns the infinite
number of
resonances which are expected to contribute already at leading order.
One may hope that the idea of {\it local cancellation} will help in
the development of a simple picture at high energies which might get
patched together with the present one. Is the simple high energy
theory a kind of string model ?

{}From a practical standpoint (without worrying about all the
theoretical issues involved in making a comparison with the
$\displaystyle{1/N_c}$
expansion) we have demonstrated that it is possible to understand
$\pi\pi$ scattering up to the $1~GeV$ region by shoehorning together
poles and contact term contributions employing a suitable
regularization procedure. It seems likely that any crossing symmetric
approximation will have a similar form. This is in the spirit of {\it
mean field} theories.

\begin{center}
{\bf Acknowledgments}
\end{center}
This work was supported in part by the U.S. DOE Contract No.
DE-FG-02-85ER40231.

\newpage
\section*{Appendix A}
\setcounter{equation}{0}
\renewcommand{\theequation}{A.\arabic{equation}}

\subsection*{Scattering kinematics}
The general partial wave scattering matrix for the multi channel case
can be written as:

\begin{equation}
S_{ab}=\delta_{ab}+2iT_{ab}\ .
\label{scatt}
\end{equation}
For simplicity, the diagonal isospin and angular momentum labels have
not been indicated.

By requiring the unitarity condition $S^{\dagger}S=1$ one deduces for
the two channel case the
following relations:
\begin{eqnarray}
Im ({T}_{11}) &=&|{T}_{11}|^2 + |{T}_{21}|^2 \nonumber \ ,\\
Im ({T}_{22}) &=&|{T}_{22}|^2 + |{T}_{12}|^2 \ , \\
Im ({T}_{12}) &=&{T}_{11}^*~{T}_{12} +
{T}_{12}^*{T}_{22}\ , \nonumber
\label{constr}
\end{eqnarray}
where ${T}_{12}={T}_{21}$. In the present case we will
identify 1 as the $\pi\pi$ channel and 2 as the $K\overline{K}$ channel.
In order to get the relations between the relative phase shifts and the
amplitude we need to consider the following parameterization of the
scattering amplitude:
\begin{equation}
S= \pmatrix
    {&\eta~e^{2i\delta_{\pi}}&\pm i\sqrt{1-\eta^2}~e^{i\delta_{\pi K}} \cr
     &\pm i\sqrt{1-\eta^2}~ e^{i\delta_{\pi K}} & \eta ~e^{2i\delta_K}
}\ ,
\label{param}
\end{equation}
where $\delta_{\pi K}=\delta_\pi +\delta_K$ and $0< \eta < 1$ is the elasticity
parameter.  By comparing
eq.~(\ref{param}) and eq.~(\ref{scatt}) one can easily deduce:
\begin{equation}
\eta^2=1 - 4|{T}_{12}|^2\ .
\label{eta2}
\end{equation}
Analogously, for ${T}_{aa}$ we have:
\begin{equation}
{T^I_{aa;l}}(s)=\frac{(\eta^I_l (s)~e^{2i\delta^I_{a;l}(s)}-1)}{2i}\ ,
\end{equation}
where $l$ and $I$ label the angular momentum and isospin, respectively.
Extracting the real
and imaginary parts via
\begin{eqnarray}
R^I_{aa;l}&=&\frac{\eta^I_l ~\sin(2\delta^I_{a;l})}{2}\ , \nonumber\\
I^I_{aa;l}&=&\frac{1-\eta^I_l ~\cos(2\delta^I_{a;l})}{2}
\label{real-imaginary}
\end{eqnarray}
\noindent
leads to the very important bounds
\begin{equation}
\big|R^I_{aa;l}\big{|}\leq\frac{1}{2}\ ,~~~~~~~0\leq I^I_{aa;l}\leq 1\ .
\label{eq:bound}
\end{equation}
The unitarity also requires $|T^I_{12;l}|< 1/2$\ .

Now we relate these partial wave amplitudes to the invariant amplitudes.
The invariant
amplitude for $ \pi_i(p_1) + \pi_j(p_2) \rightarrow  \pi_k(p_3) + \pi_l(p_4)
$ is
decomposed as:
\begin{equation}
 \delta_{ij}\delta_{kl} A(s,t,u) + \delta_{ik}\delta_{jl} A(t,s,u)
+ \delta_{il}\delta_{jk} A(u,t,s)\ ,
\label{eq:def}
\end{equation}
where $s$, $t$ and $u$ are the usual Mandelstam variables.
Note that the phase of eq.~(\ref{eq:def}) corresponds to
simply  taking
the matrix element of the Lagrangian density of a four point contact
interaction.
   Projecting out amplitudes of definite isospin yields:
\begin{eqnarray}
T_{11}^0(s,t,u) &=& 3A(s,t,u)+A(t,s,u)+A(u,t,s)\ ,\nonumber\\
T_{11}^1(s,t,u) &=& A(t,s,u)-A(u,t,s)\ ,\nonumber\\
T_{11}^2(s,t,u) &=& A(t,s,u)+A(u,t,s)\ .
\label{eq:isospin}
\end{eqnarray}

\noindent
The needed $I=0$ $\pi\pi \rightarrow K\overline{K}$ amplitude can be gotten as:

\begin{equation}
T_{12}^0(s,t,u) = -\sqrt{6}A(\pi^0(p_1)\pi^0(p_2),K^+(p_3)K^-(p_4))\ .
\end{equation}
\noindent
We then
define
the partial wave isospin amplitudes  according to the following formula:
\begin{equation}
T_{ab;l}^{I}(s)\equiv \frac{1}{2} \sqrt{\rho_{a}\rho_{b}}~\int^{1}_{-1}d\cos
\theta
P_l(\cos \theta) T_{ab}^I(s,t,u)\ ,
\label{eq:wave}
\end{equation}
where $\theta$ is the scattering angle and
\begin{equation}
\rho_a=\frac{1}{S~16\pi}\sqrt{\frac{s-4 m_{\pi}^2}{s}}~\theta(s-4 m_a^2)\ .
\label{kfact}
\end{equation}
$S$ is a symmetry factor which is 2 for identical particles
($\pi\pi$ case) and $1$ for distinguishable particles ($K\overline{K}$
case).

\newpage
\section*{Appendix B}
\setcounter{equation}{0}
\renewcommand{\theequation}{B.\arabic{equation}}
\subsection*{Chiral Lagrangian}
In the low energy physics of hadrons,
it is important to take account of
the spontaneous chiral symmetry breaking structure.
We start here with the
U(3)$_{\rm L}$$\times$U(3)$_{\rm R}$ $/$ U(3)$_{\rm V}$
non-linear realization of chiral symmetry.
The basic quantity is a 3 $\times$ 3 matrix $U$,
which transforms as
\begin{equation}
U \rightarrow U_{\rm L} U U_{\rm R}^{\dag} \ ,
\label{trans: U}
\end{equation}
where $U_{\rm L,R} \in \mbox{U(3)}_{\rm L,R}$.
This $U$ is parameterized by the pseudoscalar  $\phi$ as
\begin{equation}
U = \xi^2 \ , \qquad \xi = e^{2 i\phi/F_\pi} \ ,
\end{equation}
where $F_\pi$ is a pion decay constant.
Under the chiral transformation eq.~(\ref{trans: U}),
$\xi$ transforms non-linearly:
\begin{equation}
\xi \rightarrow
U_{\rm L} \, \xi \, K^{\dag}(\phi,U_{\rm L},U_{\rm R}) =
K(\phi,U_{\rm L},U_{\rm R}) \, \xi \, U_{\rm R}^{\dag} \ .
\end{equation}
The vector meson nonet $\rho_\mu$ is introduced as a
{\it gauge field} \cite{Kaymakcalan-Schechter}
which transforms as
\begin{equation}
\rho_\mu \rightarrow K \rho_\mu K^{\dag} +
\frac{i}{\widetilde{g}} K \partial_\mu K^{\dag} \ ,
\end{equation}
where $\widetilde{g}$ is a {\it gauge coupling constant}.
(For an alternative approach see, for a review,
Ref.~\cite{Bando-Kugo-Yamawaki:PRep}.)
It is convenient to define
\begin{eqnarray}
p_\mu &=& \frac{i}{2}
\left(
  \xi \partial_\mu \xi^{\dag} - \xi^{\dag} \partial_\mu \xi
\right) \ ,
\nonumber\\
v_\mu &=& \frac{i}{2}
\left(
  \xi \partial_\mu \xi^{\dag} + \xi^{\dag} \partial_\mu \xi
\right) \ ,
\end{eqnarray}
which transform as
\begin{eqnarray}
p_\mu &\rightarrow& K p_\mu K^{\dag} \ , \nonumber\\
v_\mu &\rightarrow& K v_\mu K^{\dag} + i K \partial_\mu K^{\dag} \ .
\end{eqnarray}
Using the above quantities
we construct the chiral Lagrangian including both pseudoscalar and
vector mesons:
\begin{equation}
{\cal L} =
-\frac{1}{2} m_v^2 \mbox{Tr}
\left[ \left( \widetilde{g}\rho_\mu - v_\mu \right)^2 \right]
- \frac{F_\pi^2}{2} \mbox{Tr}
\left[ p_\mu p_\mu \right]
-\frac{1}{4} \mbox{Tr}
\left[ F_{\mu\nu}(\rho) F_{\mu\nu}(\rho) \right],
\label{Lag: sym}
\end{equation}
where
$F_{\mu\nu} = \partial_\mu \rho_\nu -
\partial_\nu \rho_\mu - i \widetilde{g}
[ \rho_\mu , \rho_\nu ]$
is a {\it gauge field strength} of vector mesons.

In the real world chiral symmetry is explicitly broken by the quark mass term
$- \widehat{m} \overline{q} {\cal M} q$,
where $\widehat{m} \equiv (m_u+m_d)/2$,
and ${\cal M}$ is the dimension-less matrix:
\begin{equation}
{\cal M} = \left(
\begin{array}{ccc}
1+y & & \\ & 1-y & \\ & & x
\end{array} \right) \ .
\end{equation}
Here $x$ and $y$ are the quark mass ratios:
\begin{equation}
x = \frac{m_s}{\widehat{m}} \ , \qquad
y = \frac{1}{2}
\left(\frac{m_d-m_u}{\widehat{m}}\right) \ .
\end{equation}
These quark masses lead to mass terms for pseudoscalar mesons.
Moreover,
in considering the processes related to the kaon,
(in this paper we will consider
$\pi\pi\rightarrow K \overline{K}$ scattering amplitude.)
we need to take account of the large splitting of the $s$ quark mass from
the $u$ and $d$ quark masses.
These effects are included as
SU(3) symmetry breaking terms in the above Lagrangian,
which are summarized, for example, in
Refs.~\cite{Schechter-Subbaraman-Weigel,Harada-Schechter}.
Here we write the lowest order pseudoscalar mass term only:
\begin{equation}
{\cal L}_{\phi{\rm-mass}} =
\delta' \mbox{Tr}
\left[ {\cal M} U^{\dag} + {\cal M}^{\dag} U \right] \ ,
\label{pi mass 1}
\end{equation}
where $\delta'$ is an arbitrary constant.

We next introduce higher resonances into our Lagrangian.
First, we write the interaction between the
scalar nonet field $S$ and pseudoscalar mesons.
Under the chiral transformation,
this $S$ transforms as
$S \rightarrow K S K^{\dag}$.
A possible form which includes the minimum number of  derivatives is
proportional to $
\mbox{Tr}\left[ S p_\mu p_\mu \right] \ .$
The coupling of a physical isosinglet field to two pions is then
described by
\begin{equation}
{\cal L_{\sigma}}=-\frac{\gamma_0}{\sqrt{2}}\; \sigma \;
\partial_{\mu}\vec{\pi}\cdot\partial_{\mu}\vec{\pi}\ .
\label{la:sigma}
\end{equation}
Here we should note that the
chiral symmetry requires derivative-type interactions
between scalar fields and pseudoscalar mesons.
Second, we represent the tensor nonet field by
$T_{\mu\nu}$
(satisfying $T_{\mu\nu} = T_{\nu\mu}$, and $T_{\mu\mu}= 0$.),
which transforms as
$T_{\mu\nu} \rightarrow K T_{\mu\nu} K^{\dag}$.
The interaction term is given by
\begin{equation}
{\cal L}_T = - \gamma_2 F_\pi^2
\mbox{Tr}\left[ T_{\mu\nu} p_\mu p_\nu \right] \ .
\end{equation}
The heavier vector resonances such as $\rho(1450)$ can be introduced
in the same way as $\rho$ in eq.~(\ref{Lag: sym}).

\section*{Appendix C}
\setcounter{equation}{0}
\renewcommand{\theequation}{C.\arabic{equation}}
\subsection*{Unregularized amplitudes}
\subsection*{Amplitudes for the $\pi\pi \rightarrow \pi \pi$ channel}
The current algebra contribution to $A(s,t,u)$ is
\begin{equation}
A_{ca}(s,t,u)=2\frac{(s-m_{\pi}^2)}{F_{\pi}^2}\ .
\label{current}
\end{equation}
The amplitude for the vectors can be expressed in the following form
\begin{equation}
A_{\rho}(s,t,u)=-\frac{g^2_{\rho\pi\pi}}{2m^2_{\rho}}\left[
\frac{t(u-s)}{m^2_{\rho}-t}+\frac{u(t-s)}{m^2_{\rho}-u}\right]\ ,
\label{vector}
\end{equation}
where $g_{\rho\pi\pi}$ is the coupling of the vector to two pions.

For the scalar particle we deduce
\begin{equation}
A_{f_0}(s,t,u)=\frac{\gamma_0^2}{2}\frac{\left(s-2m_{\pi}^2\right)^2}
{m_{f_0}^2-s}\ .
\label{scalar}
\end{equation}

To calculate the tensor exchange diagram we need the spin 2 propagator
\cite{tensor}
\begin{equation}
\frac{-i}{m^2_{f_2}+q^2}\left[
\frac{1}{2}\left(
\theta_{\mu_1\nu_1} \theta_{\mu_2\nu_2}+
\theta_{\mu_1\nu_2}\theta_{\mu_2\nu_1}\right)-
\frac{1}{3}\theta_{\mu_1\mu_2}\theta_{\nu_1\nu_2}\right]\ ,
\label{eq:tensorpropag}
\end{equation}
\noindent
where
\begin{equation}
\theta_{\mu\nu}=\delta_{\mu\nu}+\frac{q_\mu q_\nu}{m^2_{f_2}}\ .
\end{equation}
\noindent
A straightforward computation then yields the $f_2$ contribution to the
$\pi\pi$ scattering amplitude:
\begin{eqnarray}
A_{f_2}(s,t,u)&=&\frac{\gamma^2_2}{2(m^2_{f_2}-s)}
\left(
-\frac{16}{3}m_{\pi}^4
+\frac{10}{3}m_{\pi}^2 s
-\frac{1}{3}s^2
+\frac{1}{2}(t^2+u^2)\right.\nonumber\\
&~&\left.-\frac{2}{3}\frac{m_{\pi}^2s^2}{m^2_{f_2}}
-\frac{s^3}{6m^2_{f_2}}
+\frac{s^4}{6m^4_{f_2}}
\right)\ .
\label{eq:tensorampl}
\end{eqnarray}

\subsection*{Amplitudes for $\pi^0\pi^0 \rightarrow K^+ K^-$ }

Current algebra amplitude:
\begin{eqnarray}
A_{ca}(\pi^0\pi^0,K^+K^-)&=&\frac{s}{2F_{\pi}^2}\ .
\label{12_current}
\end{eqnarray}
\noindent
Vector meson contribution:
\begin{eqnarray}
A_{Vector}(\pi^0\pi^0,K^+K^-)&=&\frac{g^2_{K^* K \pi}}{8 m^2_{K^*}}
\left[\frac{t(s-u)}{m^2_{K^*}-t} +\frac{u(s-t)}{m^2_{K^*}-u}
\right. \nonumber \\
&+&\left.
(m_k^2-m_{\pi}^2)^2\left(\frac{1}{m^2_{K^*}-t}+\frac{1}{m^2_{K^*}-u}\right)
\right]\ .
\end{eqnarray}
\noindent
Direct channel contribution for the scalar:
\begin{eqnarray}
A_{f_0}(\pi^0\pi^0,K^+K^-)&=&\frac{1}{4}{\gamma_{f_0\pi\pi}\gamma_{f_0K\overline{K}}}
\frac{(s-2m_{\pi}^2)(s-2m_k^2)}{m_{f_0}^2-s}\ .
\end{eqnarray}
\noindent
Cross channel contribution for the scalar:
\begin{eqnarray}
A_{K^*_0}(\pi^0\pi^0,K^+K^-)&=&\frac{\gamma_{K^*_0 K\pi}^2}{8}
\left[\frac{(m_K^2+m_{\pi}^2-t)^2}{m^2_{K^*_0}-t}+
\frac{(m_K^2+m_{\pi}^2-u)^2}{m^2_{K^*_0}-u}\right]\ .
\end{eqnarray}
Direct channel tensor contribution:
\begin{eqnarray}
A_{f_2}(\pi^0\pi^0,K^+K^-)&=&\frac{\gamma_{2\pi\pi}\gamma_{2K\overline{K}}}
{2(m_{f_2}^2 -s) }\left[ \left( \frac{s^2}{4m_{f_2}^2}+\frac{t}{2}-
\frac{(m_{\pi}^2+m_{K}^2)}{2}\right)^2\right.
\nonumber\\
&+&\left.
\left(\frac{s^2}{4m_{f_2}^2}+\frac{u}{2}-\frac{(m_{\pi}^2+m_{K}^2)}{2}
\right)^2\right.
\nonumber \\
&-&\left.
\frac{2}{3}\left(\frac{s^2}{4m_{f_2}^2}-\frac{s}{2}+m_{\pi}^2
\right)\left(\frac{s^2}{4m_{f_2}^2}-\frac{s}{2}+m_{K}^2
\right)\right]\ .
\end{eqnarray}
\noindent
Cross channel tensor contribution:
\begin{eqnarray}
A_{K^*_2}(\pi^0\pi^0,K^+K^-)&=&
\frac{\gamma_{2K\pi}^2}
{16(m_{K^*_2}^2 -t) }
\left\{\left[(2m_{\pi}^2-s)-\frac{1}{2m_{K^*_2}^2}(m_{\pi}^2-m_{K}^2+t)^2
\right]  \right. \nonumber \\
&\times& \left.
\left[(2m_{K}^2-s)-\frac{1}{2m_{K^*_2}^2}(m_{K}^2-m_{\pi}^2+t)^2
\right] \right. \nonumber \\
&+&\left. \left[(u-m_{\pi}^2-m_{K}^2) + \frac{1}{2m_{K^*_2}^2}(t^2-
(m_{K}^2 - m_{\pi}^2)^2) \right]^2 \right. \nonumber \\
&-&\left. \frac{2}{3}
\left[(t-m_{\pi}^2-m_{K}^2) - \frac{1}{2m_{K^*_2}^2}(t^2-
(m_{K}^2 - m_{\pi}^2)^2) \right]^2 \right\} \nonumber \\
&+&( t \longleftrightarrow u )\ .
\end{eqnarray}

\end{document}